\documentclass[12pt,onecolumn]{article}
\usepackage{amsmath}
\usepackage{amsbsy}
\usepackage[dvips]{graphicx}
\begin{document}
\def\ter{\theta_{\rm er}}
\def\si{\sin \theta}
\def\Ro{R_{\odot}}
\def\co{\cos \theta}
\def\diff{$\pa \Omega/\pa r$}
\newcommand{\Bf}{{\bf B}}
\newcommand{\ep}{{\bf e}_\phi}
\newcommand{\vf}{{\bf v}}
\newcommand{\pa}{\partial}
\newcommand{\Rs}{R_{\odot}}
\newcommand{\er}{\mbox{erf}}
\newcommand{\Bc}{B_c}
\title{Full--Sphere Simulations of a Circulation--Dominated Solar Dynamo: Exploring the Parity Issue}
\author{Piyali Chatterjee$^{1}$, Dibyendu Nandy$^{2}$, Arnab Rai Choudhuri$^{3}$}
\maketitle

{$^{1, 3}$Department of Physics, Indian Institute of Science, Bangalore-560012, India. \\email: piyali@physics.iisc.ernet.in, arnab@physics.iisc
.ernet.in\\
$^{2}$Department of Physics, Montana State University, Bozeman, MT 59717, USA.\\ email: nandi@mithra.physics.montana.edu}
\section*{Abstract}


We explore a two-dimensional kinematic solar dynamo model 
in a full sphere, based on the
helioseismically determined solar rotation profile and with an
$\alpha$ effect concentrated near the solar surface, which captures 
the Babcock--Leighton
idea that the poloidal field is created from the decay of tilted
bipolar active regions.  The meridional
circulation, assumed to penetrate slightly below the tachocline,
plays an important role. Some doubts have
recently been raised regarding the ability of such a model 
to reproduce solar-like
dipolar parity.  We specifically address the parity issue
and show that the dipolar mode is preferred when certain reasonable conditions are satisfied, the most important condition being the
requirement that the poloidal field should diffuse efficiently
to get coupled across the equator. Our model is shown to reproduce
various aspects of observational data, including the phase relation
between sunspots and the weak, diffuse field. 

\section{Introduction}

Ever since Parker (1955) formulated the turbulent dynamo theory,
varieties of very different dynamo models for the Sun have 
appeared in the literature.
Within the last few years, however, a concensus view seems to be
emerging as to what should be the basic characteristics of a
solar dynamo model.    The history of how this concensus
view emerged is discussed in the Introduction of 
Nandy \& Choudhuri (2001), 
with citations to important original papers.  A more detailed
account of the history can be found in the review by Choudhuri (2003a). 
We, therefore,
begin here by summarizing the main aspects of this concensus view
without getting into the details of history again.
 
One important ingredient of a solar dynamo model is the differential
rotation, which has now been mapped by helioseismology. 
The toroidal magnetic field must be produced by the
stretching of poloidal field lines primarily within the tachocline---the
region of concentrated vertical shear at the base of the solar convection
zone (SCZ).  The toroidal field produced in the tachocline would then rise
from there due to magnetic buoyancy to produce active regions.  Simulations
of flux rise through the SCZ suggested that the toroidal
field at the bottom must be of order $10^5$ G (Choudhuri \& Gilman
1987; Choudhuri 1989; D'Silva \& Choudhuri 1993; Fan et al.\ 1993; D'Silva \&
Howard 1993; Caligari et al. 1995).  Since such a strong field is expected to 
quench the usual mean field $\alpha$ effect (Parker 1955; Steenbeck et al.\
1966), one possibility being considered by many researchers is that
the poloidal field is generated at the surface
from the decay of tilted bipolar region---an idea that goes back to
Babcock (1961) and Leighton (1969).  The poloidal component generated
at the surface is first advected poleward by a meridional circulation
(Wang et al.\ 1989a, 1989b; Dikpati \& Choudhuri 1994, 1995; Choudhuri
\& Dikpati 1999).  Finally,
the poloidal component sinks with the downward flow at the poles and
is brought to the tachocline where it can be stretched by the 
differential rotation to generate the toroidal field---thus completing
the full cycle.  Since advection by the meridional circulation
plays such a crucial role in such a model, we would refer to this
model as the circulation-dominated solar dynamo or CDSD model.

Although the above view of the solar dynamo arose by assimilating
the ideas of many researchers over the years, Wang et al. (1991),
Choudhuri et al.\ (1995) and Durney (1995, 1996, 1997) were amongst the first
to demonstrate the crucial role which meridional circulation is expected
to play in modern solar dynamo models.  While Choudhuri et al.\ (1995)
modeled the generation of poloidal field from the decay of active regions
by introducing a phenomenological $\alpha$ parameter concentrated 
at the surface, Durney (1995, 1996, 1997) followed 
Leighton (1969) more closely to capture
this effect by taking two flux rings of opposite polarity at the surface.
Afterwards, Nandy \& Choudhuri (2001) have demonstrated that these two
approaches give qualitatively similar results.  We shall develop our models
by prescribing an $\alpha$ coefficient concentrated near the surface.
Dikpati \& Charbonneau (1999) and K\"uker et al.\ (2001) presented
CDSD models with solar-like internal rotation and with 
such a specification of $\alpha$ effect.  Since \diff\ has a larger amplitude
at the high latitudes within the tachocline (where it is negative) rather
than at low latitudes (where it is positive), solar-like rotation tends
to produce strong toroidal fields at high latitudes, in apparent contradiction 
to the observed fact that sunspots always appear at low latitudes.  Commenting
on this problem, Dikpati \& Charbonneau (1999) write: ``this is an unavoidable
inductive effect \ldots and no change in model parameters can do away
entirely with this feature''.  K\"uker et al.\ (2001) point out: ``All
recent dynamo models with the observed rotation law are faced with this
problem, even when the $\alpha$ effect has been strongly reduced in the 
polar region by the relation $\alpha \propto \sin^2 \theta \co$, as we
also did''.
 
Nandy \& Choudhuri (2002) have recently shown in a brief 
communication that this problem can be
solved by postulating a meridional flow penetrating somewhat deeper
than hitherto believed. If the meridional flow goes below the tachocline
near the poles, then the strong toroidal field produced within the
tachocline at high latitudes
is immediately pushed underneath into the convectively
stable layers and cannot emerge at the high latitudes. The meridional
flow then carries this toroidal field through the stable layers to low
latitudes.  There the meridional flow rises and the toroidal flux
enters the SCZ to become buoyantly unstable and produce
active regions at low latitudes.  Such a penetrating meridional flow
produces theoretical butterfly diagrams in remarkable agreement with the 
observed butterfly diagram.  The conventional wisdom was that the toroidal
field which forms sunspots at low latitudes must have been produced
at the low latitude.  With the helioseismically determined rotation
profile, this seems unlikely and it may well be
that the toroidal field is actually generated at the high latitude,
even though it is not allowed to erupt there.  

Recently it has been pointed out by Dikpati \& Gilman (2001) and
Bonanno et al.\ (2002) that the CDSD model with the Babcock--Leighton
mechanism for producing the poloidal field near the surface may
not give the magnetic configuration with the observed parity.
Hale's polarity law of bipolar sunspots suggests that the toroidal
magnetic field is anti-symmetric across the solar equator,
implying a dipolar parity.  A majority
of the CDSD models were solved within one hemisphere and the boundary
conditions at the equator were taken such that the dipolar mode
was forced.  Dikpati and Gilman (2001) solved
the dynamo problem in the full sphere and found that the
CDSD model with the $\alpha$ effect concentrated near the top of the
SCZ preferentially excites the quadrupolar mode in which the toroidal
field is symmetric across the equator---opposite to what is observed.
Only when the $\alpha$ effect was concentrated near the bottom of
the SCZ, they found the dipolar parity in
conformity with observations.  Bonanno et al.\ (2002) confirmed these
findings.   

The aim of the present paper is to provide the technical details of our
dynamo models not given in the earlier brief paper of Nandy \& Choudhuri (2002),
to check the parity of these dynamo models by extending our code from a
hemisphere to a full sphere and to address some related issues.  
In the dipolar mode, the poloidal field lines have to connect across
the equator.  It is necessary for the poloidal field to diffuse efficiently
for this to be possible.  On the other hand, the diffusion of the toroidal
field has to be suppressed if we want to ensure that the toroidal field
has opposite signs on the two sides of the equator.  Since the toroidal
component is much stronger than the poloidal component, we expect
the turbulent diffusion to be much less effective on the toroidal
component than on the poloidal component.  On using a high diffusivity
for the poloidal field and a low diffusivity for the toroidal field,
we find that the dipolar parity is preferred.  There is no need to
include an additional $\alpha$ effect at the bottom of SCZ to ensure
dipolar parity.  
   
Since the nature of the meridional circulation plays such a crucial
role in our model, let us make a few comments on it.
Within the last few years, helioseismic techniques have given
us some information about the sub-surface meridional circulation
to a depth of about 15\% of the solar radius (Giles et al.\ 1997;
Braun \& Fan 1999). So far there is no convincing observational
evidence for an equatorward counter-flow deeper down, though it
must exist to conserve mass.  It is generally believed that the
turbulent stresses in the SCZ drive the meridional
circulation, although the details of how this happens are not
understood (see, for example, Gilman 1986, \S3.4.2). So the 
meridional circulation is expected to be confined to the convection
zone.  However, it is conceivable that an equatorward meridional
transport of material takes place in the overshoot layer below
the bottom of SCZ.  This view is supported by recent simulations
of solar convection (Miesch et al.\ 2000).  A dynamo model with
a meridional flow through the convectively stable overshoot
layer seems, at the present time, to be 
the model with minimal extraneous assumptions which gives 
satisfactory results.  
  
In the next section, we describe the basic features of our
model.  Then \S3 focuses on the parity question and discusses
the conditions to be satisfied to ensure the correct parity.
In \S4 we present some more details of what we consider our
standard solar dynamo model, along with 
comparisons with observations.  Finally
our conclusions are summarized in \S6.    

\section{Mathematical Formulation}
\subsection{The basic equations and boundary conditions}

All our calculations are done with a code for solving the
axisymmetric kinematic dynamo problem.  An axisymmetric magnetic
field can be represented in the form
\begin{equation}
{\bf B} = B (r, \theta) {\bf e}_{\phi} + \nabla \times [ A
(r, \theta) {\bf e}_{\phi}],
\end{equation}
where $B (r, \theta)$ and $A(r, \theta)$ respectively correspond
to the toroidal and poloidal components.  The standard equations
for the so-called $\alpha \omega$ dynamo problem are:
\begin{equation}
\frac{\pa A}{\pa t} + \frac{1}{s}(\vf.\nabla)(s A)
= \eta_{p} \left( \nabla^2 - \frac{1}{s^2} \right) A + \alpha B,
\end{equation} 
\begin{equation}
\frac{\pa B}{\pa t} 
+ \frac{1}{r} \left[ \frac{\pa}{\pa r}
(r v_r B) + \frac{\pa}{\pa \theta}(v_{\theta} B) \right]
= \eta_{t} \left( \nabla^2 - \frac{1}{s^2} \right) B 
+ s(\Bf_p.\nabla)\Omega + \frac{1}{r}\frac{d\eta_t}{dr}\frac{\partial{B}}{\partial{r}}
\end{equation}\\
where $s = r \sin \theta$.  Here $\vf$ is the meridional flow, $\Omega$
is the internal angular velocity of the Sun 
and $\alpha$ is the coefficient which describes the generation
of poloidal field at the solar surface from the decay of bipolar
sunspots.  We allow the turbulent diffusivities $\eta_p$ and $\eta_t$
for the poloidal and toroidal components to be different.
We describe below how we specify $\vf$, $\Omega$, $\eta_{p}$, $\eta_{t}$,
and $\alpha$.  Once these quantities are specified, we can solve
(2) and (3) to study the behaviour of the dynamo. Apart from 
the specification
of these parameters, we also include magnetic buoyancy in a way described
below.  We carry out our
calculations in a meridional slab $R_b = 0.55R_\odot < r < R_\odot$, $0 <\theta <\pi$.

The boundary conditions are as follows.
At the poles 
($\theta = 0, \pi$) we have
\begin{equation}
A = 0, \ B = 0.
\end{equation}
For a perfectly conducting solar core, the bottom boundary ($r = R_b$) 
condition should be
\begin{equation}
A = 0,\ B = 0.
\end{equation}
However, if the bottom of the integration region is taken well below
the depths to which the meridional circulation reaches (and hence
below the depths to which magnetic fields are carried), then the solutions
are rather insensitive to the bottom boundary condition.  We carried
out some calculations by changing the bottom boundary condition of
the toroidal field from $B = 0$ to 
\begin{equation}
\frac{\pa}{\pa r} (r B) = 0.
\end{equation}
The solutions
remained virtually unchanged.	
At the top ($r = R_\odot$), the toroidal field has to be zero ($B = 0$) 
and $A$ has to match smoothly to a potential field satisfying 
the free space equation 
\begin{equation}
\left( \nabla^2 - \frac{1}{r^2 \sin^2 \theta} \right) A = 0. 
\end{equation}
Dikpati and Choudhuri (1994) describe how this is done.

We have used a finer grid resolution as compared to other models existing 
in the literature, with $129 \times 129$ grid cells in the latitudinal and 
radial directions.  The algorithm used by us in developing the numerical code
is described in the Appendix of Dikpati \& Choudhuri (1994) and the Appendix
of Choudhuri
\& Konar (2002). If either the $\alpha$ coefficient has a quenching factor
or magnetic buoyancy is included to suppress the growth of the magnetic field, 
then any arbitrary initial condition either asymptotically goes to zero 
(sub-critical condition) or relaxes to a steady dynamo solution (super-critical
condition).  We now discuss how the various parameters are specified.

\subsection{Internal rotation $\Omega$}

We use the following analytic form to represent the solar internal 
rotation (Schou et al.\ 1998; Charbonneau et al.\ 1999):
\begin{equation}
\Omega(r,\theta) = \Omega_{RZ} + \frac{1}{2}\left[1 + \er \left(2\frac
{r - r_t}{d_t}\right) \right]\left[\Omega_{SCZ}(\theta) - \Omega_{RZ}\right].
\end{equation}
This analytical expression fits the results of helioseismology
fairly closely for the following values of parameters: $r_t = 0.7R_\odot$, 
$d_t = 0.05R_\odot$ 
$\Omega_{RZ}/2\pi = 432.8$ nHz, $\Omega_{SCZ}(\theta) = \Omega_{EQ} 
+ \alpha_2 cos^2(\theta) + \alpha_4 cos^4(\theta)$, with $\Omega_{EQ}/2\pi 
= 460.7$ nHz, $\alpha_2/2\pi = -62.69$ nHz and $\alpha_4/2\pi = -67.13$ nHz. 
A contour plot of $\Omega$ generated by the above expression is shown in
Fig.~\ref{fig:diffrot}, in which the tachocline is shown as a shaded region. Since
helioseismology has already determined $\Omega$, we do not have much
freedom to vary its parameters.  At the present time, however, there
exist some uncertainties as to the exact location of the tachocline
and its thickness.  These are indicated by the parameters $r_t$ and $d_t$.
Especially, it is still not completely clear how the tachocline is located
with respect to the bottom of the convection zone (whether it is completely
below the convection zone or partly inside it).  There are some indications
that the tachocline may actually have a prolate shape, such that more of
the tachocline comes within the convection zone at higher latitudes than
at the lower latitudes.  Given the many other uncertainties in the problem,
we have not taken this effect into account in our calculations.

We take the bottom of the convection zone at $r = 0.71 R_\odot$, which is
marked in Fig.~\ref{fig:diffrot}. By bottom of the convection zone,
we mean the depth at which the temperature gradient changes from being
sub-adiabatic below to super-adiabatic above.  As we point out later,
magnetic buoyancy is assumed to be operative only above the bottom of 
the convection zone.  Since the strong toroidal field is generated in the
tachocline, a tachocline below the convection zone would imply a situation
where the toroidal field is created at a location which is immune to
magnetic buoyancy.  Hence, the location of the tachocline with respect to
the base of the convection zone is of considerable importance in our problem.
As seen in Fig.~\ref{fig:diffrot}, we take part of the tachocline above the bottom of the
convection zone.  All our calculations are done with this assumed
profile of $\Omega$.  
\begin{figure}
\begin{minipage}[t]{0.45\textwidth}
\label{fig:diffrot}
\center
\includegraphics[width=7cm,height=7cm]{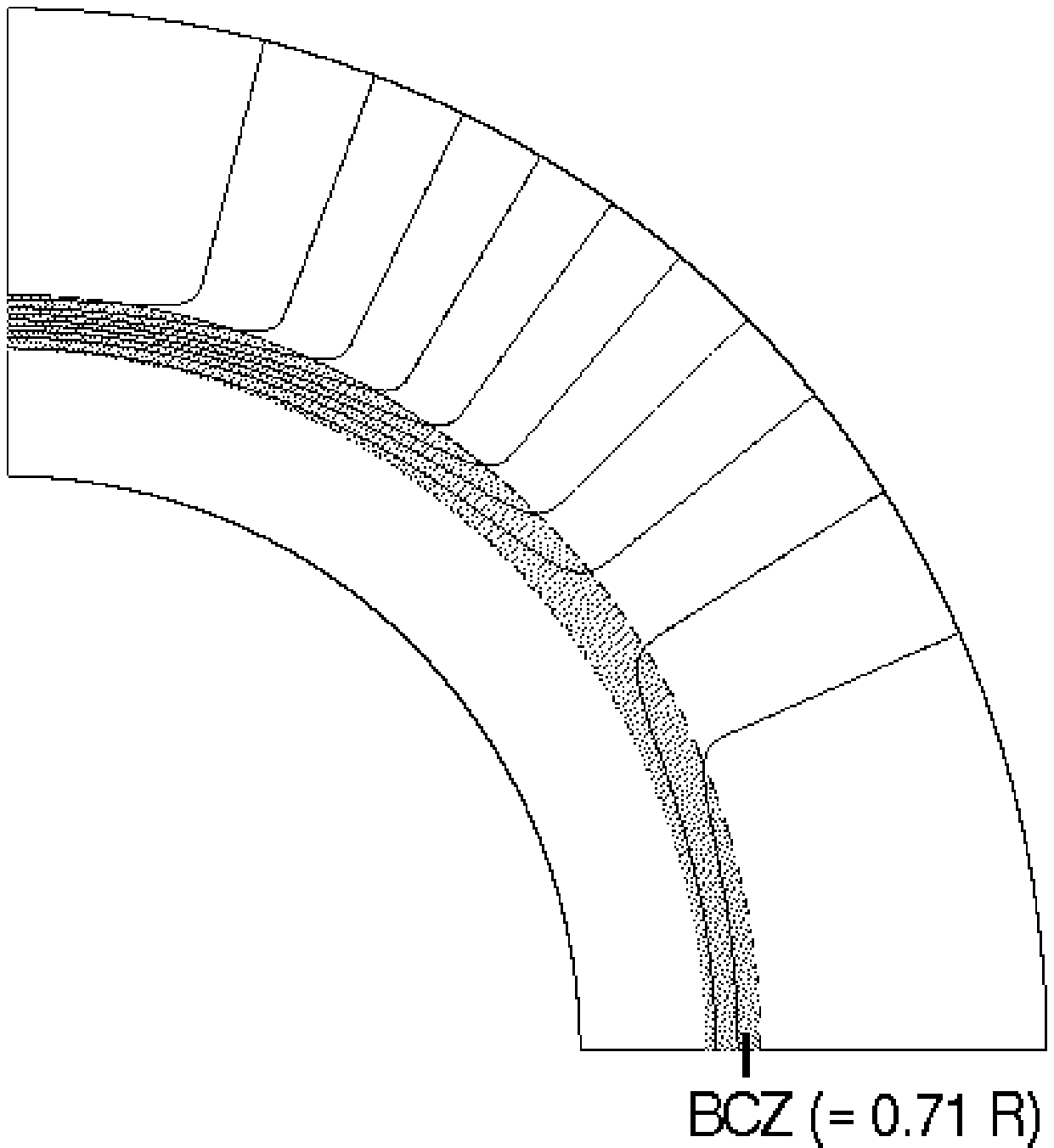}
\caption{\label{fig:diffrot}
Contours of constant angular velocity as obtained from (8)
and used in all of our calculations.  The tachocline is
indicated as the shaded region. The bottom of the convection zone (BCZ) is at 
0.71R$_{\odot}$. The differential rotation being symmetric
about the equator, only one quadrant is shown.}   
\end{minipage}%
\hspace{0.5cm}
\begin{minipage}[t]{0.45\textwidth}
\label{fig:merd}
\center
\includegraphics[width=4cm,height=7cm]{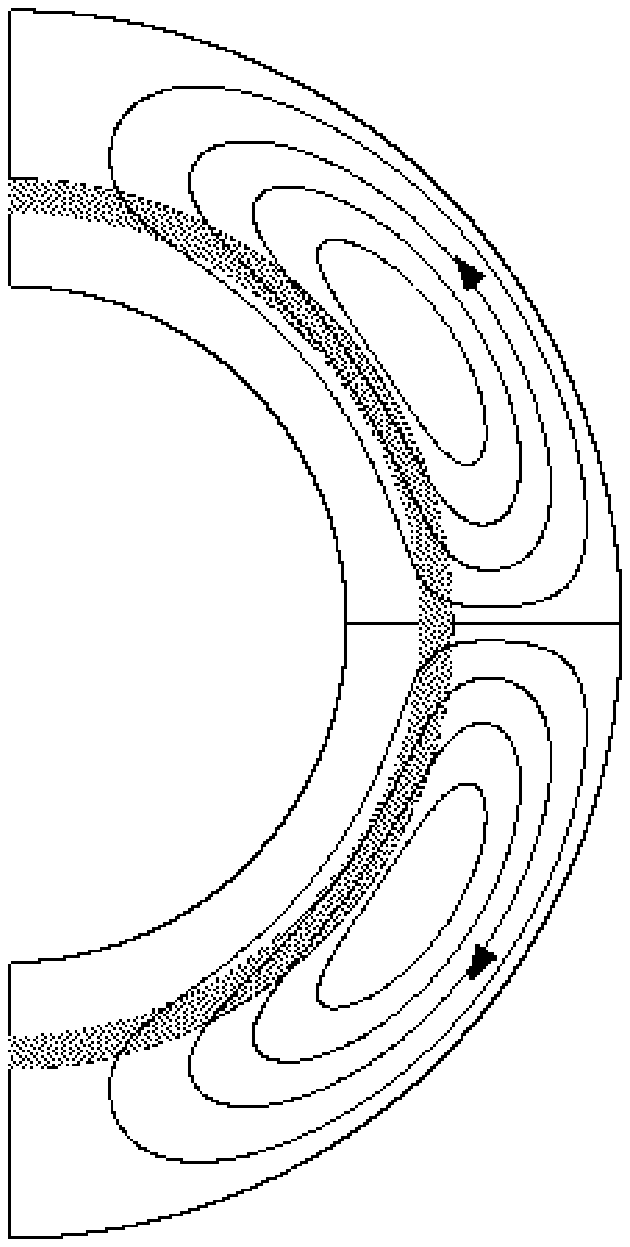}
\caption{\label{fig:merd}
Streamlines of meridional circulation
obtained by taking penetration radius $R_p = 0.61 R_\odot$.  The tachocline 
is shown as a shaded
region. Arrows denote the direction of flow.}   
\end{minipage}
\end{figure}

In some of our earlier work (Choudhuri et al.\ 1995; Nandy \& Choudhuri 2001;
Nandy 2002), we had taken $\Omega$ to be independent of $\theta$ and used
a radial variation appropriate for the equatorial region.  If $\Omega$ is
taken to be a function of $r$ alone, then the behaviour of the dynamo is
much simpler.   On the other hand, when $\Omega$ is a function of both
$r$ and $\theta$ as given by (8), the problem becomes immensely more
complicated.

\subsection{Meridional circulation \vf}

We now describe how the meridional circulation is specified in the northern
hemisphere.  The circulation in the southern hemisphere is simply obtained
by a mirror reflection of the velocity field across the equator. 
We get the meridional circulation from the stream function $\psi$
defined through the equation
\begin{equation}
\rho \vf = \nabla \times [\psi (r, \theta) {\bf e}_{\phi}].
\end{equation}
Assuming a density stratification 
\begin{equation}
\rho = C \left( \frac{R_\odot}{r} - \gamma \right)^m,
\end{equation}
we take
\begin{equation}
\psi r \sin \theta = \psi_0 (r - R_p) \sin \left[ \frac{\pi (r - R_p)}
{(\Rs - R_p)} \right] \{ 1 - e^{- \beta_1 r \theta^{\epsilon}}
\}\{1 - e^{\beta_2 r (\theta - \pi/2)} \} e^{-((r -r_0)/\Gamma)^2}
\end{equation}
\begin{figure}[t]
\label{fig:vtheta}
\center
\includegraphics[width=10.0cm,height=8cm]{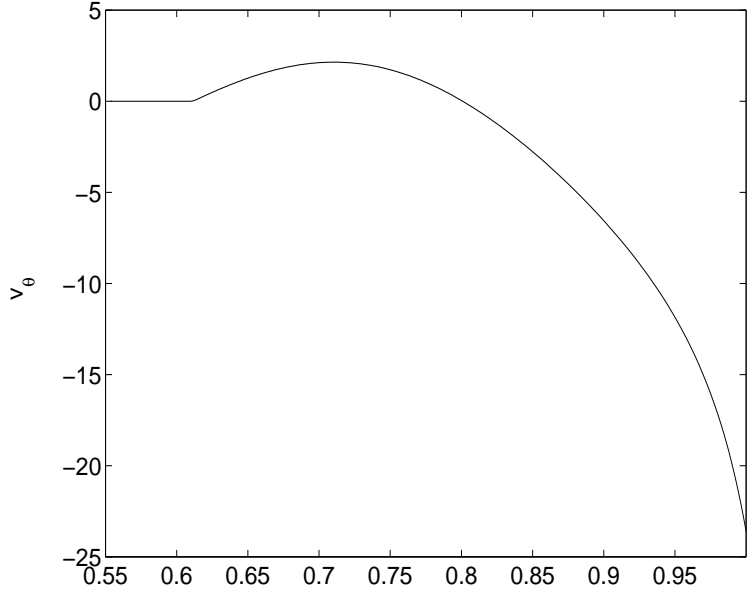}
\caption{\label{fig:vtheta}Plot of $v_{\theta}$ (in m s$^{-1}$) as a function of $r$ at the
mid-latitude $\theta = 45^{\circ}$. }   
\end{figure}
with the following values of the parameters: 
$\beta_1 = 1.36 \times 10^{-8}$ m$^{-1}$, $\beta_2 = 1.63 \times 10^{-8}$ 
m$^{-1}$, $\epsilon = 2.0000001$, $r_0 = (\Rs - R_b)/4.0$, $\Gamma = 
3.47 \times 10^{8}$ m, $\gamma = 0.95$, $m=3/2$.  It is the parameter $R_p$
which determines the depth to which the meridional circulation
penetrates. Fig.~\ref{fig:merd} shows the streamlines of the meridional circulation
obtained by taking $R_p = 0.61 R_\odot$, which is used in all our
calculations and which corresponds to the meridional flow going slightly 
below the tachocline.  The amplitude of the meridional circulation is
fixed by $\psi_0/C$.  We choose it such that the poleward flow near
the surface at mid-latitudes peaks typically around $v_0=22$ m s$^{-1}$. The equatorward 
counterflow peaks at the core-convection zone interface and has a value of 1.8 m s$^{-1}$, which
is similar to what observational analysis of sunspot drift suggests (Hathaway et al. 2003).
Since the form of the meridional circulation seems crucial for the
stability for the dynamo, let us make some comments on it.  On comparing
(11) with equation (9) of Nandy \& Choudhuri (2001), it will be found
that we have now added an extra factor $(r - R_p)$ just after $\psi_0$.
This extra factor ensures that $v_{\theta}$ smoothly falls to zero at
$R_p$.  If this factor is not included, then $v_{\theta}$ has a finite
value at $R_p$, leading to a discontinuity at $R_p$ if there is no flow
below.  Fig.~\ref{fig:vtheta} shows the profile of $v_{\theta}$ at the mid-latitude
obtained with the factor $(r - R_p)$ included. It should also be noted in
Fig.~\ref{fig:vtheta} that $v_{\theta}$ in our model decreases monotonically
below the bottom of the SCZ and becomes very small in the tachocline.  
We want to emphasize that we need very modest flows below the tachocline
(presumably not much beyond the region of overshooting) to make our model work.

We point out that we find well-behaved periodic solutions only for
certain forms of the meridional circulation.
K\"uker et al.\ (2001) also found oscillatory solutions only within
limited regions of parameter space.  However, the dynamo becomes very
robust and stable with a sufficiently deeply penetrating meridional
flow having a smooth $v_{\theta}$ profile.
   
\subsection{Diffusion coefficients $\eta_p$ and $\eta_t$}

We expect the turbulent diffusivity inside the convection zone $\eta_{SCZ}$ 
to be much larger than the diffusivity $\eta_{RZ}$ in the radiative interior,
the overshoot layer being the region within which the value of diffusivity
makes a transition from $\eta_{RZ}$ to $\eta_{SCZ}$. Since the poloidal 
component is weak, the turbulent diffusivity acts on it without any difficulty.
We take the diffusivity for the poloidal component to be
\begin{equation}
\eta_{p}(r) = \eta_{RZ} + \frac{\eta_{SCZ}}{2}\left[1 + \er \left(\frac{r - r_{BCZ}}
{d_t}\right) \right]
\label{eq:etap}
\end{equation}
Fig.\ref{fig:diff} shows a plot of $\eta_p$ with the following values of the parameters: 
$\eta_{SCZ} = 2.4 \times 10^{12}$cm$^2$ s$^{-1}$, 
$\eta_{RZ} = 2.2 \times 10^8$ cm$^2$ s$^{-1}$, $r_{BCZ} = 0.7 R_\odot$, 
$d_t = 0.05R_{\odot}$.
\begin{figure}
\begin{minipage}[t]{0.45\textwidth}
\label{fig:diff}
\centerline{\includegraphics[height=6.0cm,width=6cm]{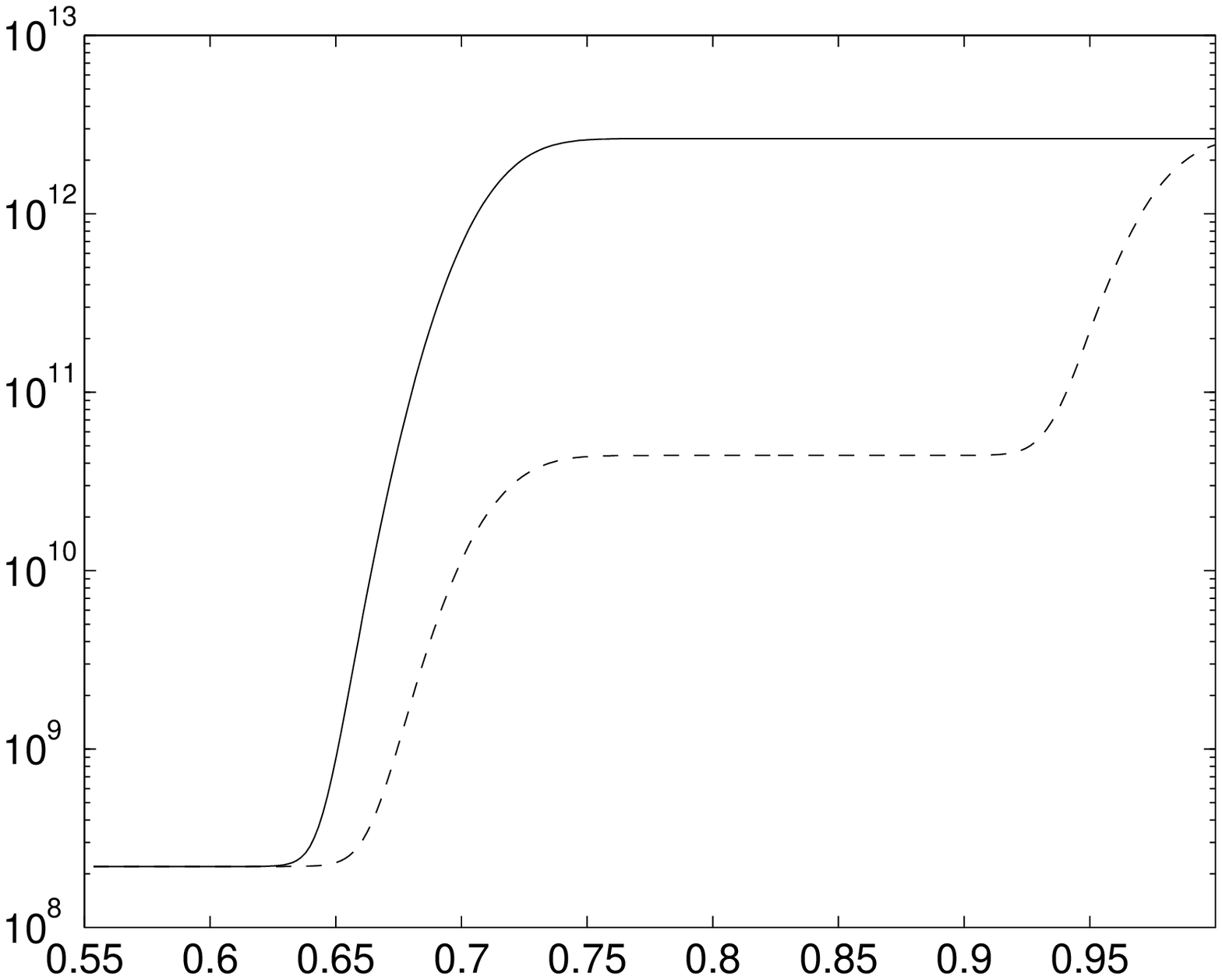}}
\caption{\label{fig:diff} Plots of $\eta_{p}(r)$(solid) and $\eta_{t}(r)$ (dashed)
as given by (\ref{eq:etap}) and (\ref{eq:etat}) as functions of the fractional radial distance ($r/R_{\odot}$). y-axis is in units of cm$^2$ s$^{-1}$.} 
\end{minipage}%
\hspace{0.5cm}
\begin{minipage}[t]{0.45\textwidth}
\label{fig:diffquad}
\centerline{\includegraphics[height=6.0cm,width=6cm]{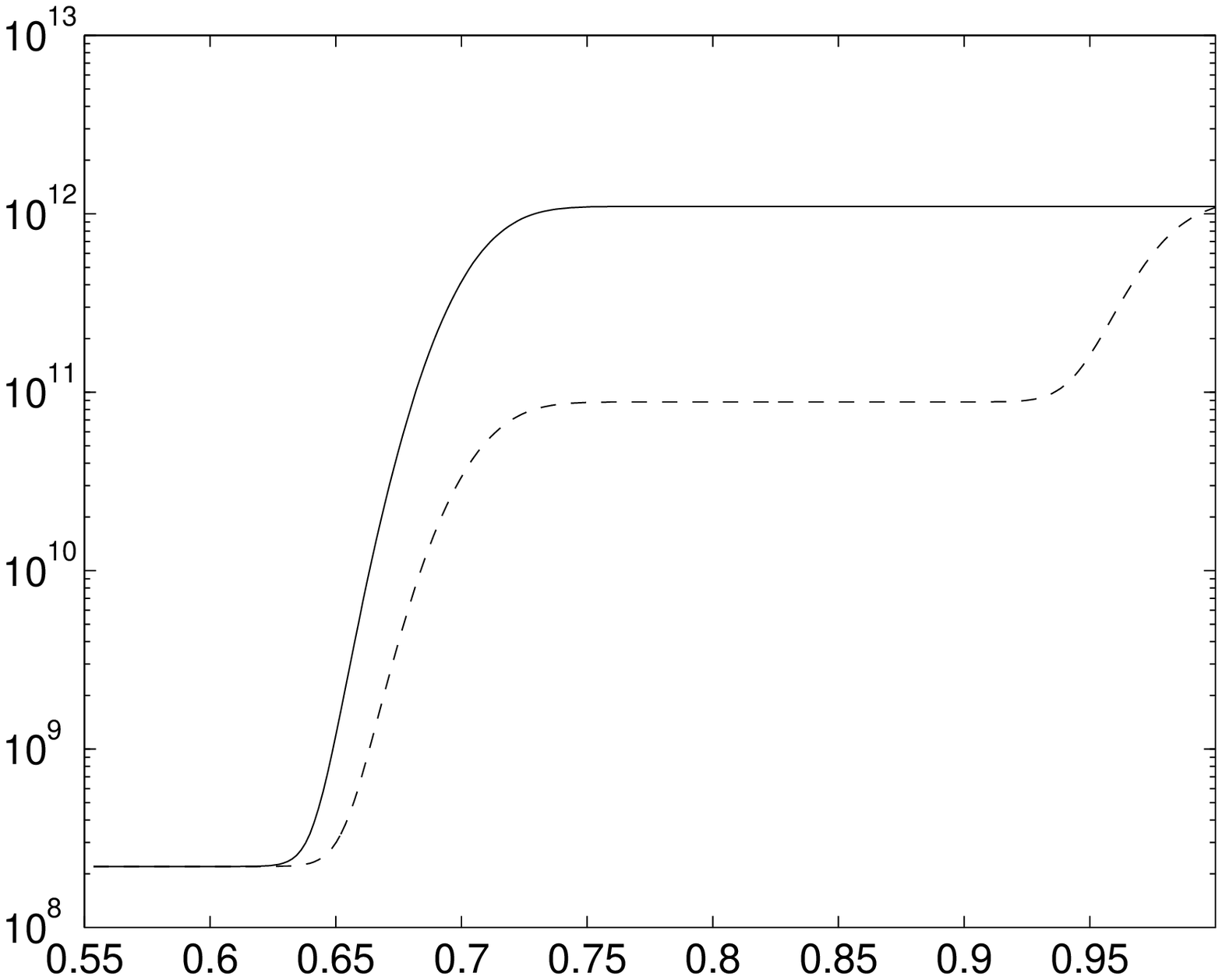}}
\caption{\label{fig:diffquad} Plots of $\eta_{p}(r)$(solid) and $\eta_{t}(r)$ (dashed) in 
cm$^2$ s$^{-1}$ used in \S 3.2, as function of the fractional radial distance ($r/R_{\odot}$).} 
\end{minipage}
\end{figure}
The toroidal component, however, has a value larger than the equipartition 
value till it rises to about 30,000 km or 40,000 km below the solar surface
(Longcope \& Choudhuri 2002, \S2).  Only in the top layers of the SCZ, the
diffusion coefficient of the toroidal component should be equal to the usual
turbulent diffusivity value $\eta_{SCZ}$.  Within the main body of SCZ, the
action of turbulent diffusivity on the toroidal component must be considerably
suppressed.  In view of this, we take the diffusivity $\eta_t$ of the toroidal
component as shown in Fig.~\ref{fig:diff}, which is generated from the expression
\begin{equation}
\label{eq:etat}
\eta_{t}(r) = \eta_{RZ} + \frac{\eta_{SCZ1}}{2}\left[1 + \er \left(\frac{r - r^{\prime}_{BCZ}}
{d_t}\right) \right] + \frac{\eta_{SCZ}}{2}\left[ 1 + \er \left(\frac{r-r_{TCZ}}{d_t}
\right) \right]
\end{equation} 
with $\eta_{SCZ}=4\times10^{10}$cm$^2$ s$^{-1}$,  
$r^{\prime}_{BCZ} = 0.72R_{\odot}$ and $r_{TCZ}=0.95 R_{\odot}$.  As we shall
see below, $\eta_p$ and $\eta_t$ specified in this way gives solutions with
dipolar parity.  Except \S3.2--3, everywhere else in our paper we use
$\eta_p$ and $\eta_t$ as specified above.

 Simulations of the 
evolution of the weak, diffuse field on the solar surface (Wang et al. 1989a,b), 
as well as observational estimates from sunspot 
decay, point out that $\eta_{SCZ}$ must be of the order of $10^{12}$ cm$^2$ 
s$^{-1}$ in the upper layers of the convection zone.  We have taken a
value of this order on the higher side and find that it allows sufficient
diffusion of the poloidal component across the equator to enforce the
dipolar mode.  
A low value of diffusivity below the bottom of the convection zone is very
important in dynamo models with meridional flow penetrating below the
tachocline.  A low diffusivity in the tachocline and the overshoot
layer ensures that the toroidal field which is produced
in the high latitudes within the tachocline does not decay much while
being transported to low latitudes by the meridional flow (Nandy 2002).
The assumed value of $\eta_{RZ}$ essentially ensures that the magnetic
field is frozen for time scales of the order of dynamo period.  By taking
a low $\eta_t$ at the bottom of SCZ, we also
make sure that there is not much
cross-diffusion of the toroidal component
across the equator and it is possible for the toroidal
component to have opposite values in the two hemispheres. It may be noted
that there is a term involving $d \eta_t/dr$ in the evolution equation (3)
for the toroidal component.  This term has the form of an advection term,
with $d \eta_t/dr$ corresponding to a downward velocity.  We discovered an
error in the original code which produced the results of Nandy \& Choudhuri
(2002).  The error made this term involving $d \eta_t/dr$ somewhat smaller
than what it should have been.  However, on incorporating the 
suppression of $\eta_t$ in the body of the SCZ as we do here,
the gradient $d \eta_t/dr$ is moved to the upper layers (which can be seen
from Fig.~\ref{fig:diff}) and the results essentially remain the
same as earlier.

For the sake of comparison, we present in \S3.2--3 some results obtained with
a lower diffusivity for the poloidal field.  The profile of this $\eta$
is shown in Fig.~\ref{fig:diffquad}.   This low
diffusivity does not allow the poloidal components in the two hemispheres
to connect across the equator and usually the quadrupolar mode is preferred,
as we shall see in \S3.2--3.

\subsection{The $\alpha$ coefficient}

The $\alpha$ coefficient is taken in the form
\begin{equation} 
\alpha =\alpha_0 \cos \theta \frac{1}{4}
\left[ 1 + \er \left(\frac{r - r_1}{d_1} \right) \right]
\times
\left[ 1 - \er \left(\frac{r - r_2}{d_2} \right) \right].
\end{equation}
The parameters we use are $r_1 =0.95 \Rs$, $r_2 = \Rs$, $d_1 = d_2 =
0.025 \Rs$, making sure that the $\alpha$ effect is concentrated
in the top layer $0.95 \Rs \leq r \leq \Rs$. 
The solid line in Fig.~\ref{fig:alpha} shows the variation of $\alpha$ with $r$. The amplitude 
$\alpha_0$ is taken such that the dynamo is super-critical.  
We had taken $\alpha_0 =$ 25 m s$^{-1}$ in most of our
calculations, where the solutions are found to be super-critical for
such a value of $\alpha_0$. We use this form of $\alpha$ coefficient
in all our calculations except in \S3.3, where an additional $\alpha$
effect within the SCZ is included. The dashed line in Fig.~\ref{fig:alpha} shows
the radial profile of $\alpha$ used in \S3.3, the angular variation
being taken as $\cos \theta$ as used elsewhere in the paper.
\begin{figure}
\label{fig:alpha}
\centerline{\includegraphics[width=8cm,height=8cm]{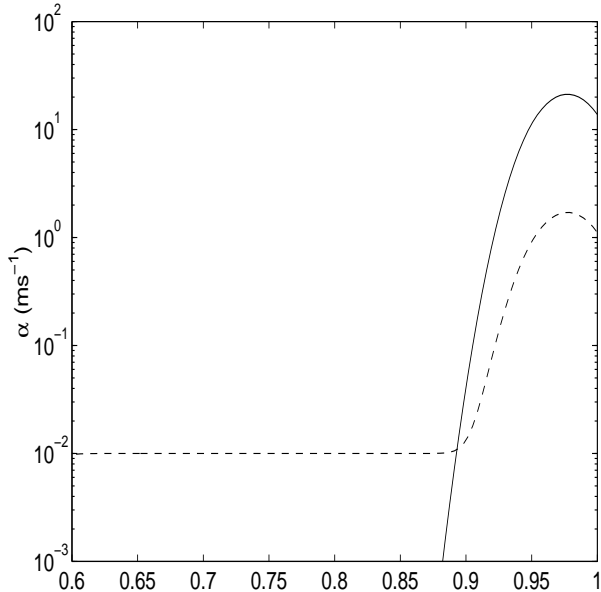}}
\caption{\label{fig:alpha} The Babcock-Leighton $\alpha$ used for most of our
calculations (solid line) and also in our standard model described in \S 3.1. The dashed line shows
the $\alpha$ used in \S 3.3, as a function of fractional radial distance $r/R_{\odot}$. 
Units are in m s$^{-1}$.}
\end{figure}
It should be noted that $\alpha$ coefficient in a Babcock--Leighton dynamo
is {\it not} given by the mean helicity of turbulence as in mean
field MHD (see, for example, Choudhuri 1998, \S16.5).  In our approach
as well in the approach of several other authors (Choudhuri et al.\ 1995;
Dikpati \& Charbonneau 1999; Nandy \& Choudhuri 2001; K\"uker et al.\ 2001),
the $\alpha$ coefficient phenomenologically captures the effect of
poloidal field generation from the decay of tilted active regions
near the solar surface. The angular factor $\cos \theta$ arises from
the angular dependence of the Coriolis force which causes the tilts of
active regions. Several groups suppressed $\alpha$ artifically at high latitudes
to reduce the strength of the toroidal field at high latitudes.
Dikpati \& Charbonneau (1999) used the angular factor $\cos \theta \sin
\theta$, whereas K\"uker et al. (2001) used $\cos \theta \sin^2 \theta$.
Flux tube simulations which calculate
the tilts of emerging active regions (D'Silva \& Choudhuri 1993; Fan
et el.\ 1993) give us some clues about the angular dependence of the
$\alpha$ coefficient.  If the 
Coriolis force could freely produce the tilt without any
opposing force operative, then the tilts at different latitudes would
have been proportional to the Coriolis factor $\cos \theta$ and the
$\alpha$ coefficient would have the same angular dependence.  However,
the Coriolis force is opposed by magnetic tension which tries to prevent
the tilt from becoming very large.  We thus find that the tilt does not
increase with latitude as fast as the Coriolis factor $\cos \theta$.
On these grounds, we expect that $\alpha$ will not increase with latitude
as fast as $\cos \theta$, but taking the angular dependence to be
$\cos \theta \sin \theta$ or $\cos \theta \sin^2 \theta$ may be unrealistic.

We also point out that we have eliminated the 
$\alpha$-quenching factor typically taken to be of the form
$(1+B^2/B_0^2)^{-1}$. If magnetic buoyancy is not included,
then such a factor is the only
source of nonlinearity in the model and  is essential to allow 
the simulation to relax to steady solutions. 
We found the $\alpha$-quenching to be redundant in presence of 
magnetic buoyancy which limits 
the toroidal field to values $< 10^{5}$ G (see \S2.5). On theoretical grounds
also, the removal of $\alpha$-quenching is quite
logical in the Babcock--Leighton dynamo models.  In mean field MHD, the 
$\alpha$ effect comes 
from helical turbulence which is quenched when the magnetic field is
super-equipartition.  In our model, however, the $\alpha$ effect is due to 
the decay of tilted bipolar 
regions.  Flux tube simulations do show that the tilt is less for
stronger magnetic fields at the bottom of the SCZ 
(see D'Silva \& Choudhuri 1993).  So the $\alpha$ effect should depend
on the magnetic field at the bottom of SCZ, but not on the local
value of the magnetic field at the surface where the $\alpha$ effect
is operative. Since our
formulation of magnetic buoyancy (see \S2.6) makes the toroidal field
erupt only when it is of order $10^5$ G, we do not expect much variation
in $\alpha$ with the magnetic field. 

\subsection{Magnetic buoyancy}

We prescribe magnetic buoyancy in a way which has
been discussed in detail by Nandy \& Choudhuri (2001) and
Nandy (2003). We search 
for toroidal field $B$ exceeding the critical field $B_c = 10^5$ G, 
{\it{above}} the base of the SCZ 
taken at $r =0.71$ at intervals of time 
$\tau = 8.8 \times 10^5$ s. Wherever $B$ exceeds $B_c$, a fraction $f = 0.5$ of 
it is made to erupt to the surface layers, with the toroidal field values 
adjusted appropriately to ensure flux conservation.  The parameter $f$ 
measures the strength of magnetic buoyancy.  It was found by Nandy \&
Choudhuri (2001) that, when $f$ was still small compared to 1 ($f$ has to
be less than 1), magnetic buoyancy already reached saturation and the
character of the dynamo did not change any more on increasing $f$.  The
value $f = 0.5$ used throughout our paper already puts the dynamo in the
buoyancy-saturated regime.

Although $A$ and $B$ in general evolve according to (2) and (3), we allow
abrupt changes in $B$ after intervals of $\tau$ to take account of
magnetic buoyancy.  While our treatment of magnetic buoyancy may not
be fully satisfactory, note that uncertainties remain in the way buoyancy
has been handled earlier by other groups, e.g.\ by treating buoyancy as a
simple loss term (Schmitt \& Sch\"ussler 1989) or by treating it in a non-local manner by making
the poloidal source term near the surface proportional to the toroidal
field strength at the bottom of the SCZ (Dikpati \& Charbonneau 1999).

\section{The parity question}

All our calculations are done with internal rotation $\Omega$, meridional
circulation ${\bf v}$ and magnetic buoyancy as specified in \S2.2, \S2.3 and
\S2.6.  We first present results in \S3.1 obtained with the diffusion
coefficients shown in Fig.~\ref{fig:diff} and an $\alpha$ effect concentrated
near the surface as shown in Fig.~\ref{fig:alpha} by the solid line. The dipolar
mode is the clearly preferred mode and the results qualitatively match the
observational data quite well.  Some authors (Dikpati \& Gilman 2001;
Bonanno et al.\ 2002) obtained anti-solar quadrupolar modes in their
calculations.  We believe that this was due to the low diffusivity of
the poloidal component which did not allow this component to connect across
the equator.  We present results in \S3.2 in which the diffusion coefficients
are as shown in Fig.~\ref{fig:diffquad}, i.e.\ the diffusivity of the poloidal component is
reduced compared to what we use in \S3.1 and that of the toroidal field is increased. 
We end up with the anti-solar quadrupolar parity in this case. Then we
show in \S3.3 that we get back the dipolar parity if we include an $\alpha$ 
effect within the SCZ (as shown by dashed line in Fig.~\ref{fig:alpha}), 
while keeping all the other things the same as in \S3.2.  This is again in
agreement with what Dikpati \& Gilman (2001) and Bonanno et al.\ (2002) found.  
We are thus able to reproduce the results of Dikpati \& Gilman (2001) and
Bonanno et al.\ (2002).  However, we do not agree with their conclusion
that a pure Babcock-Leighton $\alpha$ effect concentrated
near the solar surface cannot give the dipolar parity.  As we show in \S3.1,
the dipolar parity is the preferred parity if the diffusivity of the poloidal
component is sufficiently high, even when the $\alpha$ effect is concentrated
near the solar surface.

\subsection{Solution with dipolar parity}
\begin{figure}[h!]
\centerline{\includegraphics[height=6cm, width=10cm]{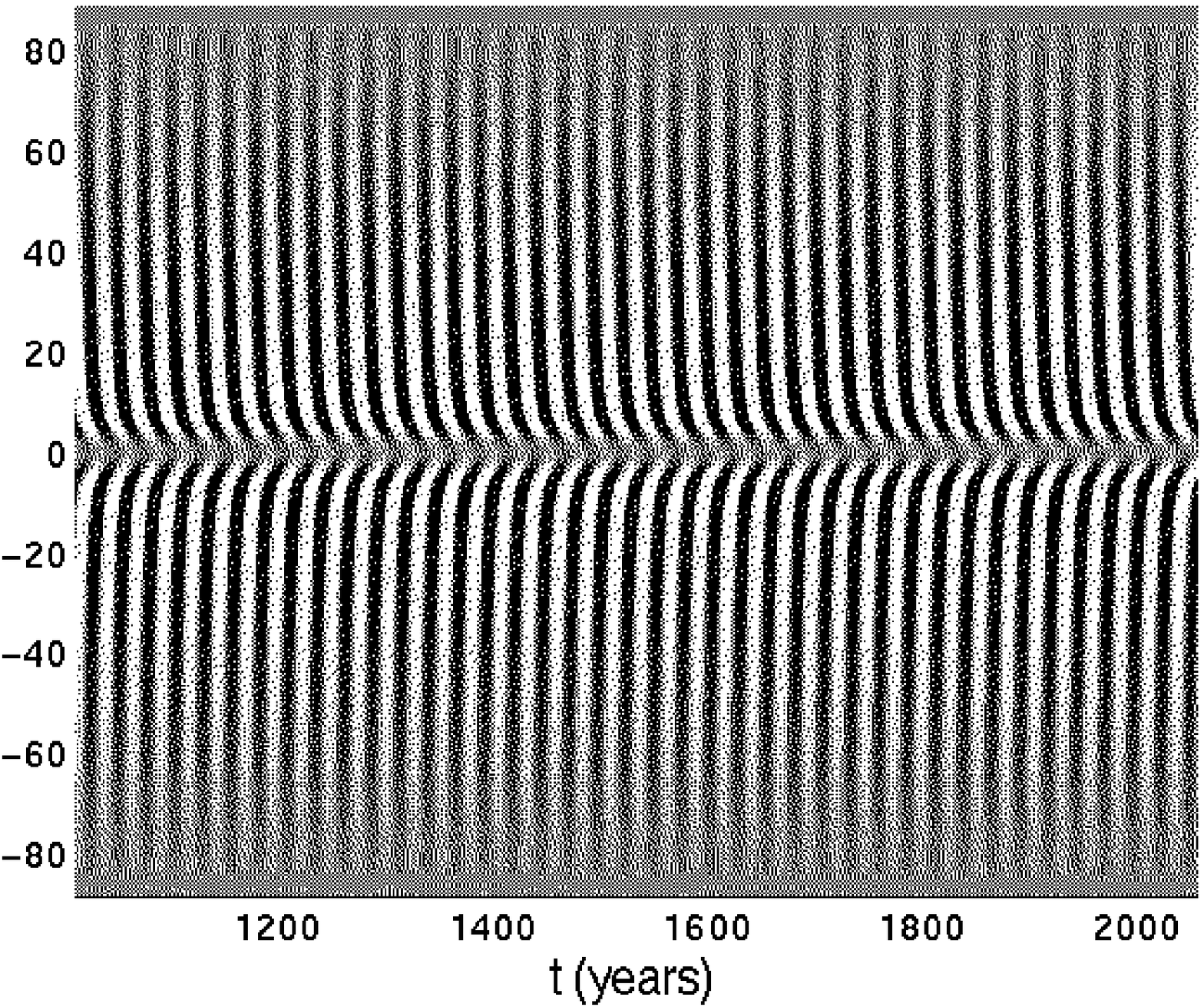}}
\vspace{0.5cm}
\label{fig:dip3000}
\centerline{\includegraphics[height=6cm, width=10.0cm]{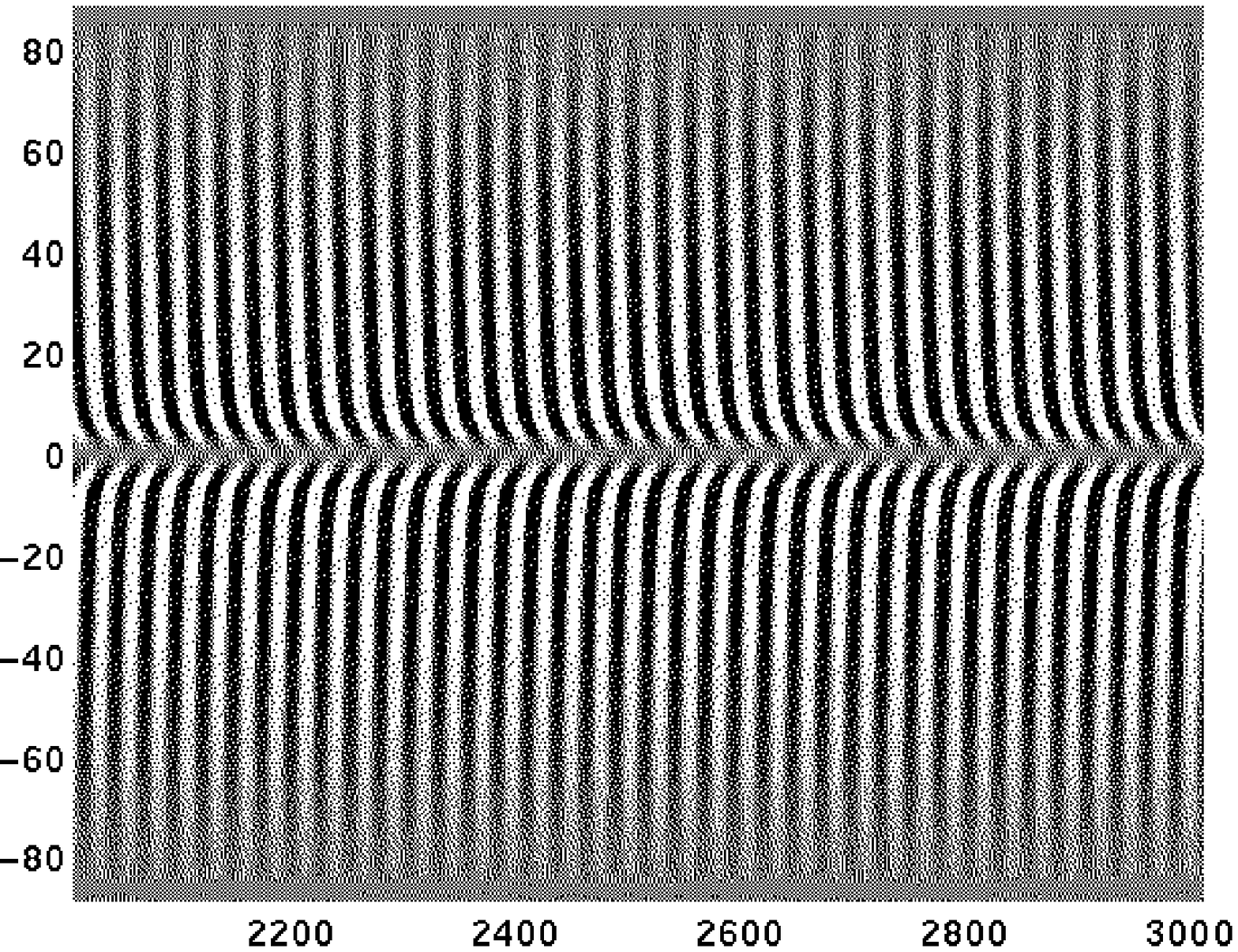}}
\caption{\label{fig:dip3000} A time-latitude plot showing the evolution of the toroidal 
field at the bottom of the 
convection zone ($r$ = 0.71 $R_{\odot}$) starting from an initial quadrupolar (anti-solar) mode to a 
pure dipolar (solar-like) mode taking a duration of 3000 years, for the case presented in \S 3.1. 
The upper panel shows the evolution from 
1000 - 2000 years whereas the lower panel traces the last thousand years (2000 - 3000 years). Regions shaded in white show positive $B$, whereas black regions denote negative $B$.}
\end{figure}

We first present a purely Babcock--Leighton dynamo ($\alpha$ effect concentrated
near the surface as shown in Fig.~\ref{fig:alpha} by solid line), 
which settles into dipolar
parity.  As we discussed above, the preferred parity depends on the diffusion
coefficients.  
Various experiments with the parameter space of the model tell us that there are two important conditions for getting the right parity.
\begin{enumerate}
\item The term $\eta_{RZ}$ representing the molecular diffusivity in the overshoot layer and 
the radiation zone below must be sufficiently small($\sim$ 2.2$\times$ 10$^{8}$ 
cm$^{2}$ s$^{-1}$) to prevent the toroidal field from diffusing across the equator.
\item The diffusivity of the poloidal field $\eta_{p}$ within the SCZ must be 
sufficiently large ($\sim$ 2.4 $\times$ 10$^{12}$ cm$^{2}$ s$^{-1}$) to allow diffusive 
coupling of the poloidal field between two hemispheres.
\end{enumerate}
\noindent Further, we have to avoid a large gradient $d \eta_t/ dr$ at the bottom of the SCZ
in order to obtain well-behaved solutions.
\begin{figure}[h!]
\label{fig:snap0}
\centerline{\includegraphics[height=8cm, width=8cm]{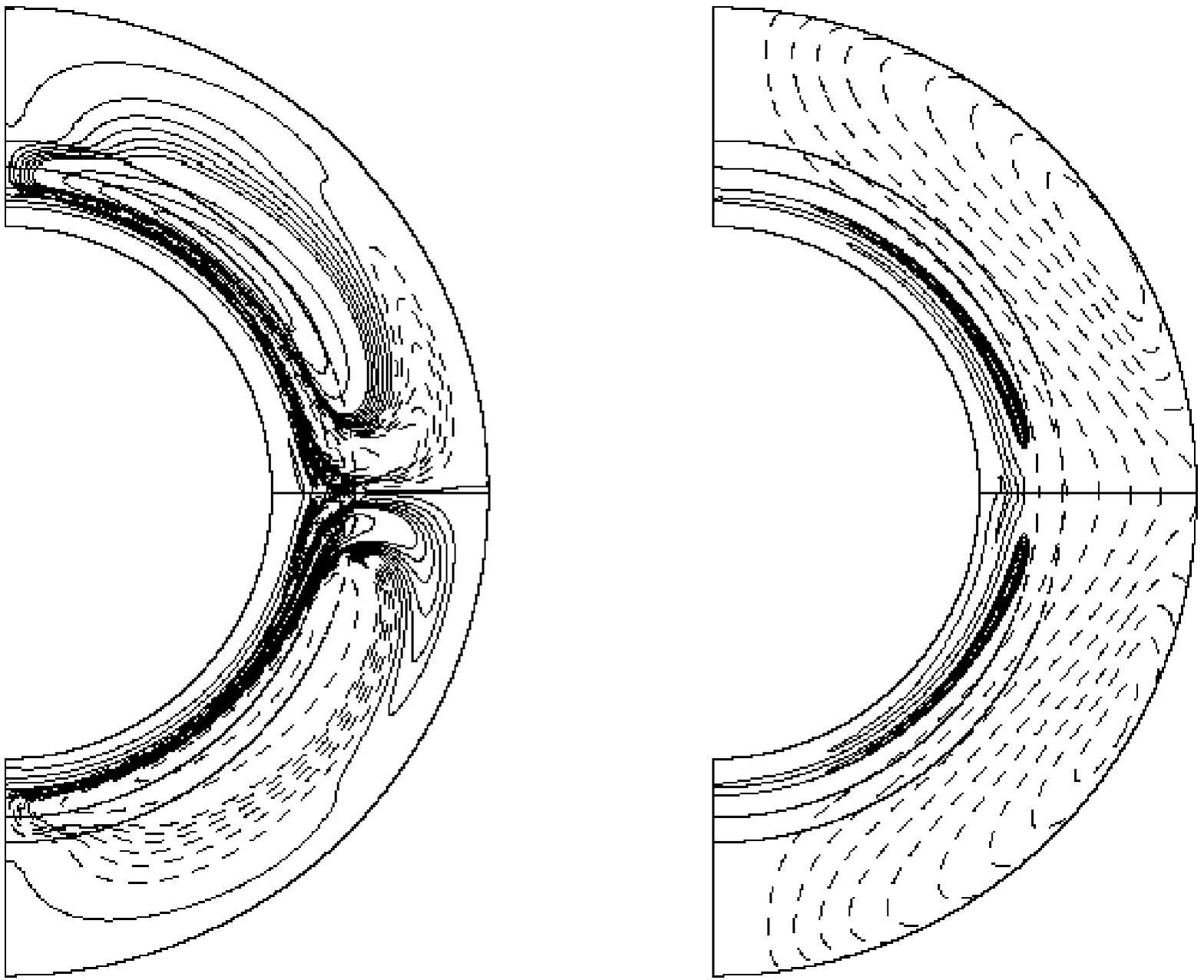}}
\caption{\label{fig:snap0} A snapshot of (a) the contours of toroidal field $B$ and 
(b) streamlines of the poloidal field, given by contours of constant $A$sin$\theta$, for 
the case of dipolar parity solution presented in \S3.1.
The solid lines denote positive $B$ or $A$, whereas dashed lines denote negative 
$B$ or $A$. }
\end{figure}
The particular solution we present here is obtained with the diffusion coefficients
as given in Fig.~\ref{fig:diff}.
To ensure that the dipolar parity is the dominant mode in the model, we start from 
a pure quadrupolar initial condition and find that the solution eventually 
relaxes to a pure dipolar parity.
Fig.~\ref{fig:dip3000} shows the
evolution of the initial quadrupolar field into a dipolar 
field within a duration of 3000 years. Fig.~\ref{fig:snap0} gives a 
snapshot of the relaxed magnetic field configuration, showing
that the poloidal field lines have connected across the equator, whereas the toroidal
field within the tachocline has opposite signs on the two sides of the equator.
We carried out the simulation for another 3000 years {\it after} the solution relaxed to the 
dipolar mode, and the solution remained in the solar-like parity state during the entire run.
We shall provide further details of this solution in \S4.
    
\subsection{Solution with quadrupolar parity}
\begin{figure}[h!]
\label{fig:quad1000}
\centerline{\includegraphics[height=10cm, width=16cm]{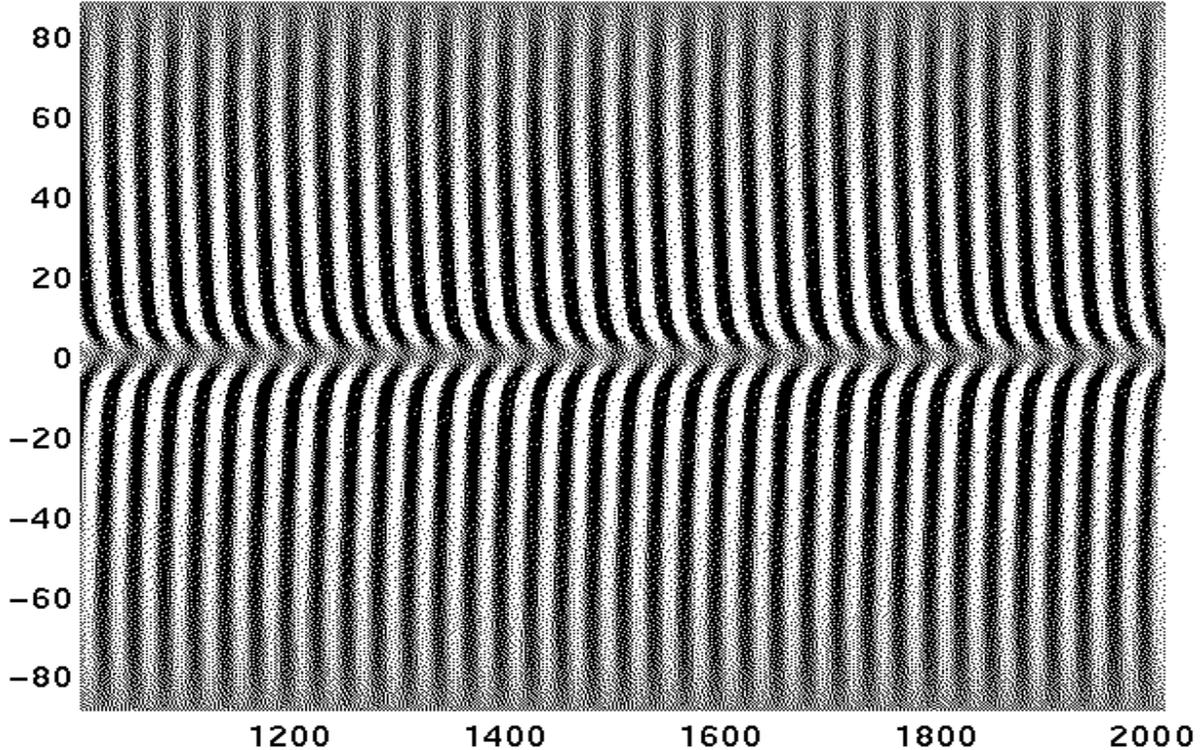}}
\caption{\label{fig:quad1000} Time-latitude plot showing the evolution of $B$ at the 
bottom of the convection zone 
from an initial dipolar state to the preferred quadrupolar 
state, taking a duration of 2000 years, for the case presented in \S 3.2. Only the last 1000 years of 
model run is shown.}
\end{figure} 
We now present a solution obtained by keeping all the parameters the same as in \S3.1,
except that we change the diffusion coefficients to what is shown in Fig.~\ref{fig:diffquad}
 rather than what is shown in Fig.~\ref{fig:diff} (as used in \S3.1).  
The lower diffusivity of the
poloidal field does not allow it to connect across the equator and we find that
the quadrupolar mode is preferred. To make sure that indeed the dominant mode in this case
is the quadrupolar mode, we started with an initial condition having dipolar parity and found
that it relaxed to a quadrupolar parity. Fig.~\ref{fig:quad1000} 
presents a theoretical butterfly diagram
showing this transition process, whereas Fig.~\ref{fig:quadsnap} illustrates the magnetic field configuration
after the dynamo has settled into a quadrupolar 
mode.  We find that the poloidal field has not diffused enough to connect across the equator,
but has remained separated on the two sides of the equator.  On the other hand, the toroidal
field in the tachocline has the same sign on the two sides of the equator and makes up
a patch of a common sign across the equator.
We have made many runs for other values of diffusion coefficients.  As we already mentioned,
a low diffusivity $\eta_{RZ}$ below the bottom of the SCZ is an
essential requirement to
obtain the dipolar parity (to ensure that the toroidal field in the tachocline cannot
diffuse much across the equator).   When we take $\eta_{RZ}$ larger than about
$2\times 10^{9}$ cm$^{2}$ s$^{-1}$, we found that we always got quadrupolar solutions
and it was not possible to get dipolar solutions even by increasing $\eta_p$ in the
SCZ to facilitate the coupling of the poloidal field across the equator.  We have shown
in Fig.~\ref{fig:quad1000} a transition from a dipolar parity to a quadrupolar parity 
in a case in which
the quadrupolar parity is preferred. How fast such a transition takes place depends on
the value of $\eta_{RZ}$. On using $\eta_{RZ} \sim  2\times10^{10}$ cm$^{2}$ s$^{-1}$, the 
change-over takes place within just 300 years, 
whereas decreasing $\eta_{RZ}$ by two orders of magnitude stretches
the time scale of transition to the quadrupolar parity to 2000 years. 

\begin{figure}[h!]
\label{fig:quadsnap}
\centerline{\includegraphics[height=8cm, width=8cm]{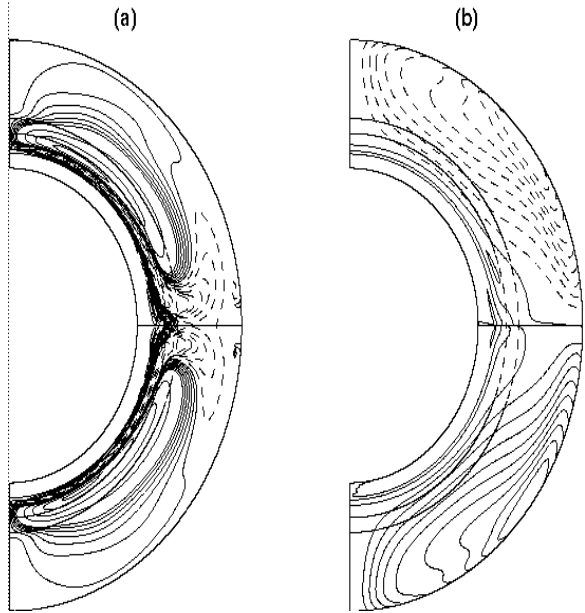}}
\caption{\label{fig:quadsnap}  A snapshot of (a) the contours of the toroidal field  and  
(b) streamlines of the poloidal field, for the quadrupolar parity solution presented in \S3.2. }
\end{figure} 
\subsection{Solution with $\alpha$ effect inside the SCZ}

Finally we consider what influence an additional $\alpha$ effect within 
the SCZ
has on the parity of the solution.  For this purpose, we change the radial 
profile
of the $\alpha$ effect to what is shown in Fig.~\ref{fig:alpha} by a dashed line
rather than the solid line in the same figure, keeping the other things the same as in 
\S3.2 (including the diffusion coefficients which are as shown in 
Fig.~\ref{fig:diffquad}).  We find that the solution relaxes to dipolar 
parity even if we start from quadrupolar parity, as shown in 
Fig.~\ref{fig:alpbfly}.  
A snapshot of the field configuration is 
presented in Fig.~\ref{fig:alpsnap}.  This is a case in which the quadrupolar parity would have 
been preferred if the additional $\alpha$ effect inside the SCZ were not 
present, as we saw in \S3.2.  However, this additional $\alpha$ effect, even though
its value inside the SCZ is much smaller than the value of the 
Babcock--Leighton $\alpha$ at the surface, can make the solution dipolar.  This is in 
agreement with what has been found by Dikpati \& Gilman (2001) and Bonanno et al.\ 
(2002).
One has to keep in mind that the $\alpha$ coefficient has to be multiplied
by the toroidal field B to provide the source term for the poloidal 
field.
Since B is very large at the bottom of the SCZ, even a very small 
$\alpha$
there can make $\alpha B$  as large as what it is near the surface.  That 
is why we find that even a very small $\alpha$ inside the SCZ or at its base 
can affect the nature of the solution so drastically. Presumably, an
$\alpha$ effect within the SCZ creates some poloidal field there
which can diffuse across the equator more efficiently than poloidal
field created near the surface, thereby enforcing the dipolar parity. Note that in the presence of
a small $\alpha$ inside the SCZ the magnitude of the Babcock-Leighton $\alpha$ required at
the surface for steady oscillating solutions decreases by almost 
a factor of 10 (dashed line in Fig.\ref{fig:alpha}). 

\begin{figure}[h!]
\label{fig:alpbfly}
\centerline{\includegraphics[width=10cm,height=6cm]{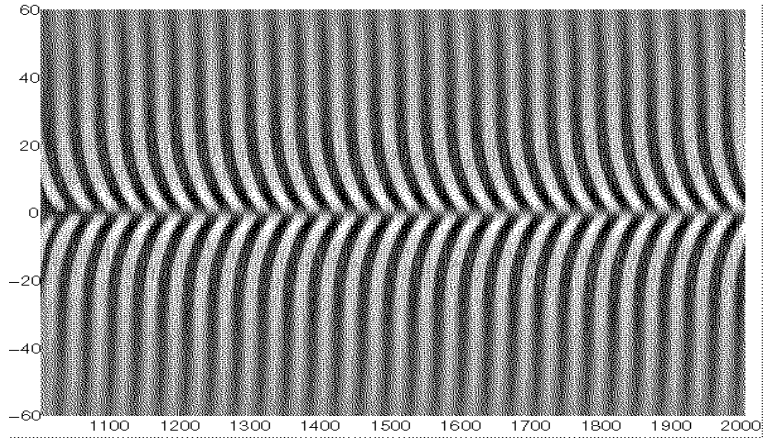}}
\caption{\label{fig:alpbfly} Time-latitude plot showing the evolution of the toroidal field $B$ 
at the bottom of the 
convection zone, with an additional $\alpha$ introduced 
inside the SCZ, as presented in \S 3.3. 
Last 1000 years of the 2000 year long model run is shown.} 
\end{figure}

Choudhuri (2003b) has argued that the strong toroidal field at the bottom 
of the SCZ would be highly intermittent and the $\alpha$ effect may be 
operative in the intervening regions of weak field.  Several other authors 
(Ferriz-Mas et al.\ 1994; Dikpati \& Gilman 2001) argue that the various instabilities
associated with the strong field also can produce something like an 
$\alpha$ effect. So an additional $\alpha$ effect within the SCZ or at
its bottom is certainly a realistic possibility.  However, our knowledge
about it at the present time is very incomplete and uncertain.  On the
other hand, we see tilted active regions decay on the solar surface and
we directly know from observations that there is an $\alpha$ effect at the
solar surface. By the Occam's razor argument, we feel that it is desirable
to first construct solar dynamo models with this Babcock--Leighton 
$\alpha$
alone---especially since we have shown in \S3.1 that it is possible to get
solar-like dipolar parity with such dynamo models.  We present a more
detailed discussion of such pure Babcock--Leighton dynamo models in \S4.

\begin{figure}[h!]
\label{fig:alpsnap}
\centerline{\includegraphics[width=10cm,height=8cm]{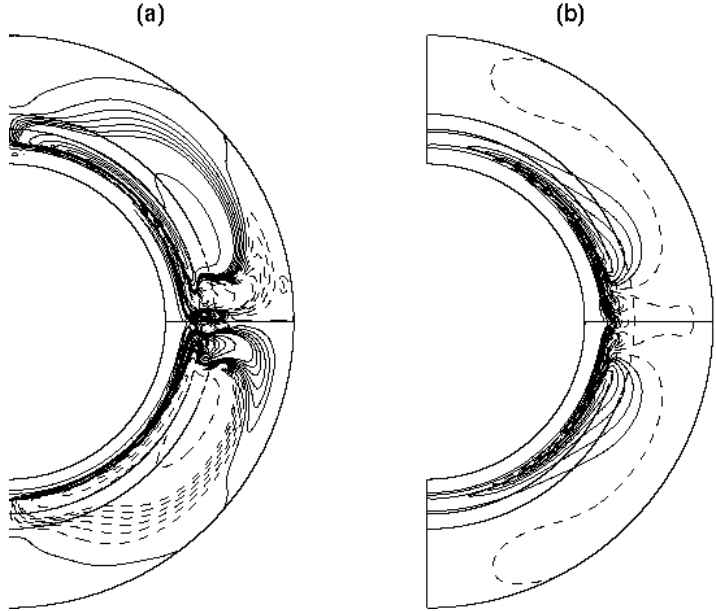}}
\caption{\label{fig:alpsnap}  A snapshots of
(a) the toroidal field contours and (b) streamlines of the poloidal 
field, for the solution presented in \S 3.3.}
\end{figure}

\section{Towards a Standard Model}

We have seen in \S3.1 that a CDSD model with an $\alpha$ effect concentrated
near the surface and with appropriate values of various parameters can give
a solution with the dipolar parity.  We would refer to this solution as our
{\em standard model}.  We have focused primarily on the parity of this solution
in \S3.1.  Now we discuss other aspects of this solution and show that this
solution matches observational data quite well.  We have already discussed
in \S2 and \S3.1 how the various parameters of this particular case are
specified.  The amplitude
of the Babcock-Leighton $\alpha$ effect used to generate our standard 
solution is 25 m s$^{-1}$. For this standard case, we refined our 
grid to have 257$\times$257 points, and the results remained unchanged with this finer resolution.

Fig.~\ref{fig:dipradbfly} shows the time-latitude contour plot of the radial field
at the solar surface, with the theoretical butterfly diagram superimposed upon it.
The butterfly diagram is producing by marking the locations of eruption, `+'
indicating the positive value of $B$ at the bottom of the SCZ which erupts and
`o' indicating the negative value. The sunspot eruptions are confined within
$\pm$ 40$^o$ and the butterfly diagrams have shapes similar to what is observationally
found (see, for example, Fig.~6.2 in Choudhuri 2003a; Hathaway
et al.\ 2003). The weak radial field
migrates poleward at higher latitudes, in conformity with observations.
One of the important aspects of observational data is the phase relation between
the sunspots and the weak diffuse field.  See \S1 of Choudhuri \& Dikpati (1999)
for a detailed discussion of this (especially note Fig.~1 there). 
The polar field
changes from positive to negative at the time of a sunspot maximum corresponding
to a negative toroidal field $B$ at the base of SCZ. This is clearly seen in the
theoretical results shown in Fig.~\ref{fig:dipradbfly}.  
In Fig.\ref{fig:dipsnap} we 
show four snapshots of the toroidal field contours and poloidal field lines 
taken successively at an interval of $1/8^{\rm{th}}$ 
the solar cycle period which happens to be about 25 years in this case. 
In all the snapshots we see that the poloidal field lines are symmetric about the equator. 
The $\eta_{p}$ in the 
convection zone is sufficiently high ($\sim 2.4 \times 10^{12}$ cm$^{2}$ s$^{-1}$) to ensure 
that the poloidal fields connect smoothly across the equator.  
On the other hand, the toroidal fields
on the two sides of the equator, which have opposite signs, 
cannot diffuse together due to the low $\eta_t$
near the base of SCZ.

\begin{figure}[h!]
\label{fig:dipradbfly}
\centerline{\includegraphics[width=10cm, height=8cm]{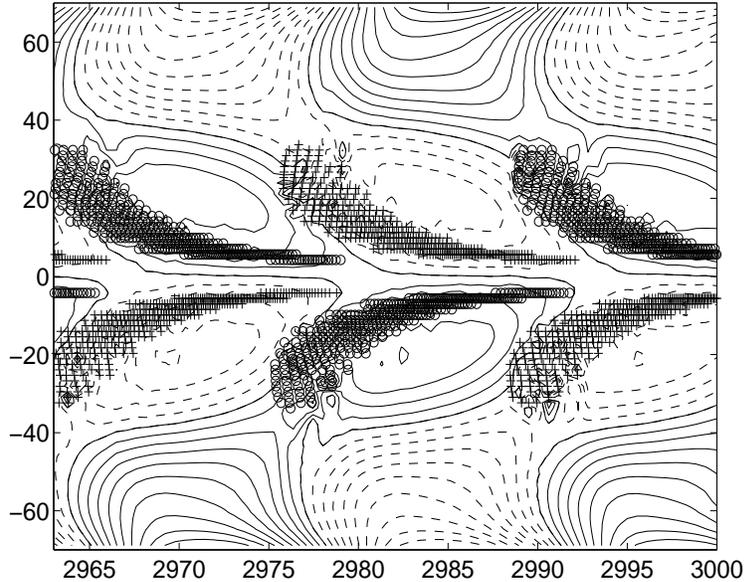}}
\caption{\label{fig:dipradbfly} Theoretical butterfly diagram of eruptions for our standard model. 
The background shows contours of diffuse radial field. Eruption latitudes are denoted by symbols 'o' and '+', indicating
negative and positive toroidal field respectively.
The dashed contours are for negative $B_{r}$, whereas  
the solid contours are for positive $B_r$. Note that negative toroidal fields give rise to negative radial field near the poles after decaying and vice versa 
in accordance with Hale's polarity law.}
\end{figure}

\begin{figure}[h!]
\label{fig:dipsnap}
\centerline{\includegraphics[width=7.0cm, height=5.0cm]{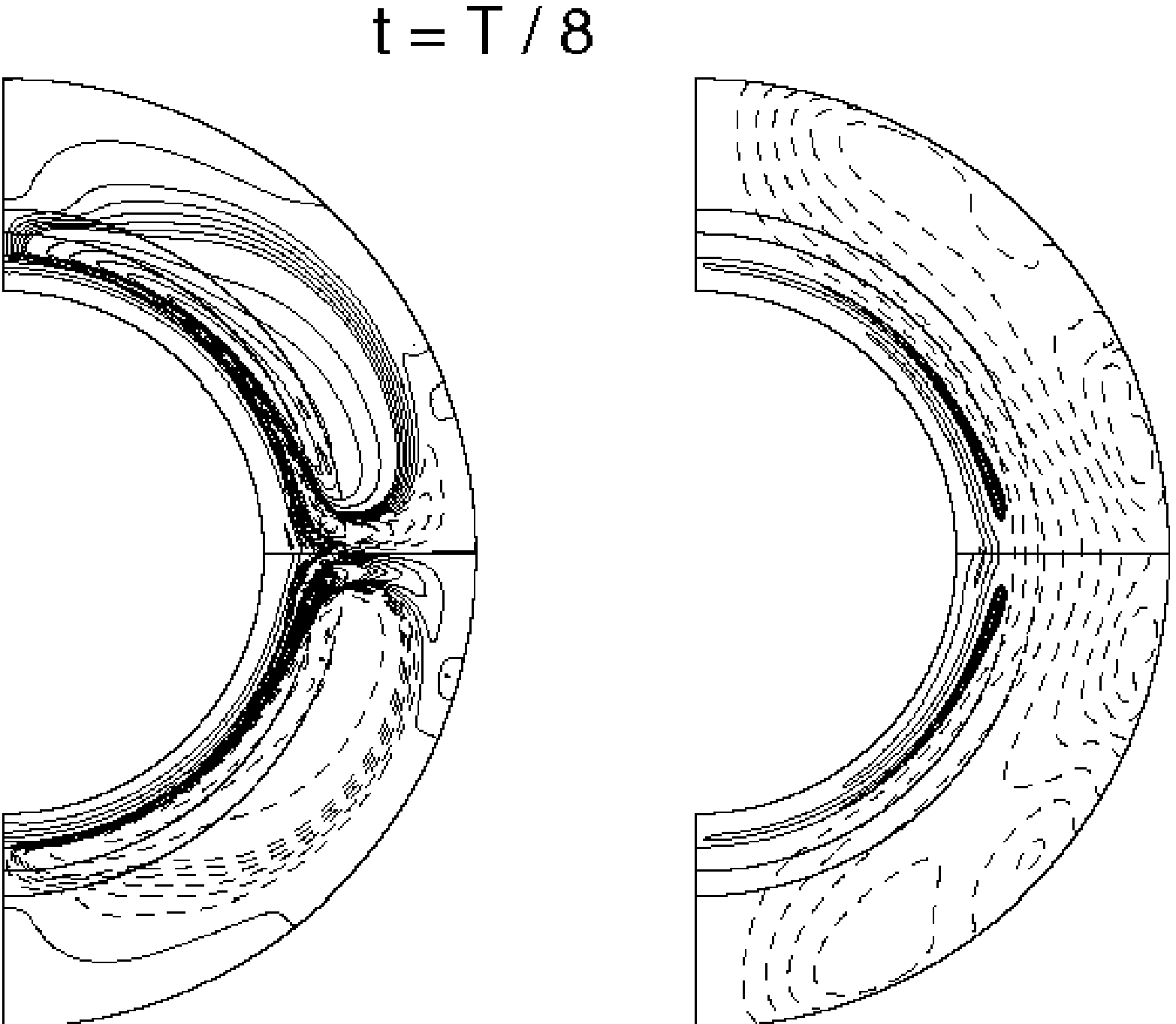}}
\centerline{\includegraphics[width=7.0cm, height=5.0cm]{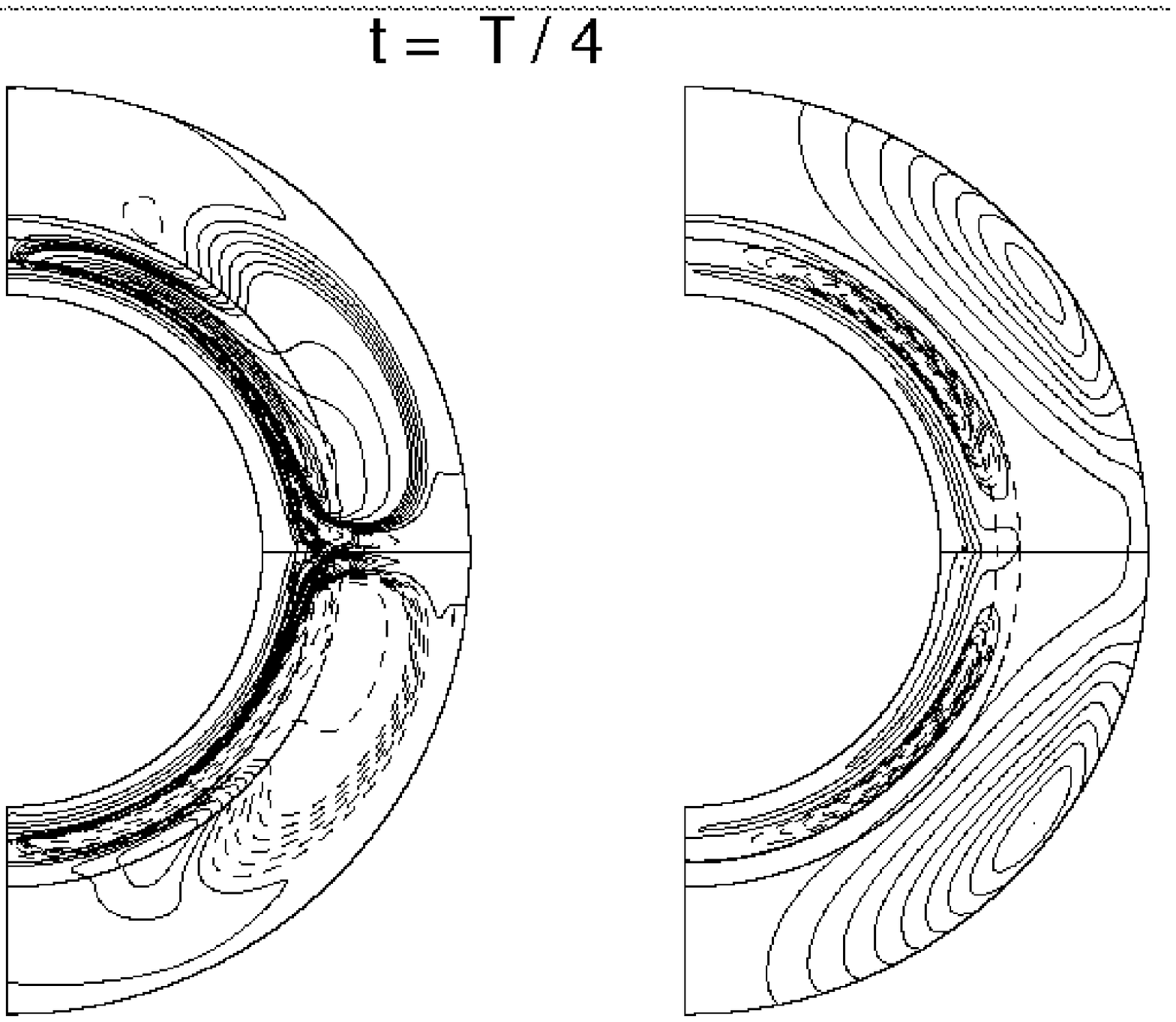}}
\centerline{\includegraphics[width=7.0cm, height=5.0cm]{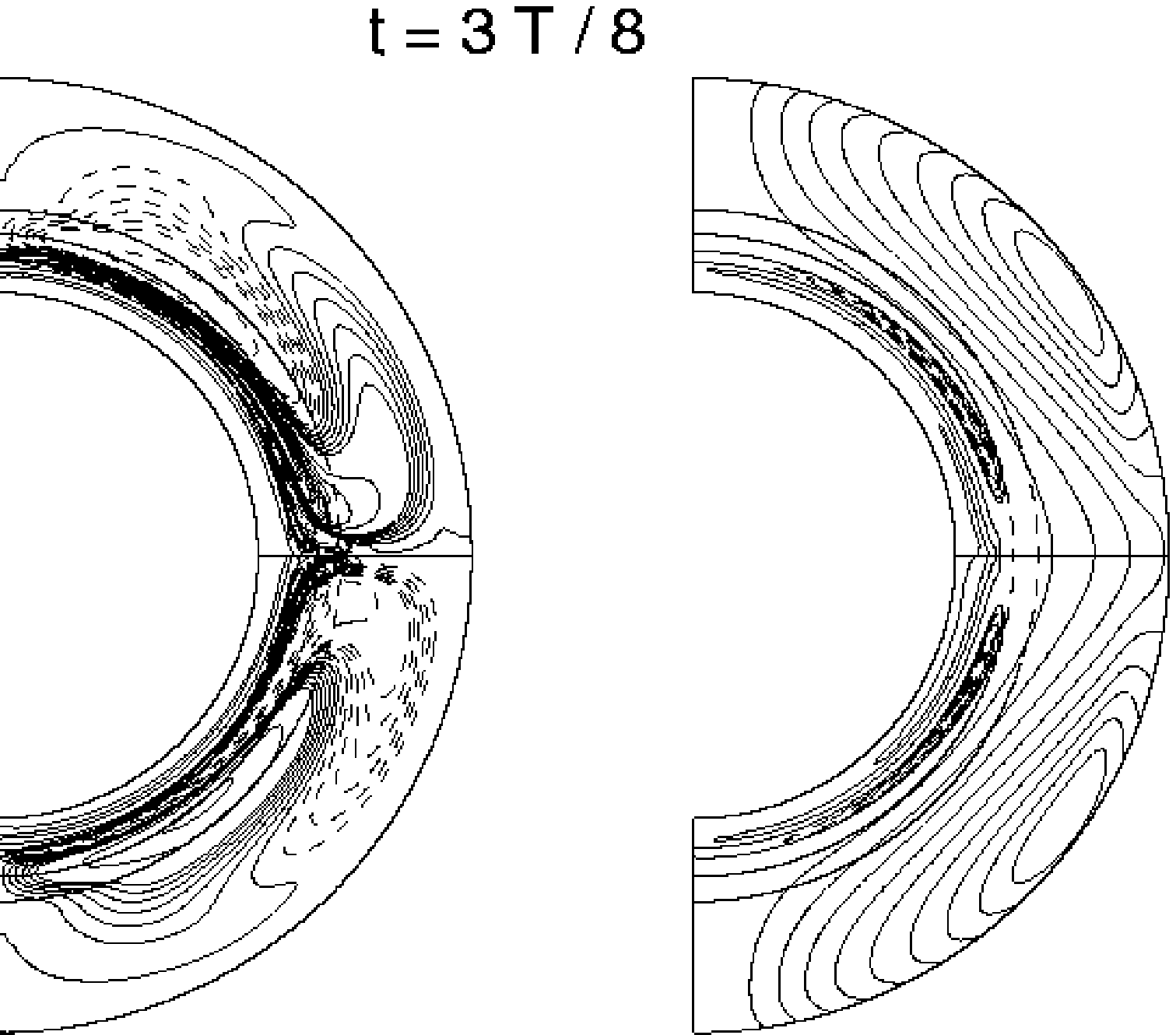}}
\centerline{\includegraphics[width=7.0cm, height=5.0cm]{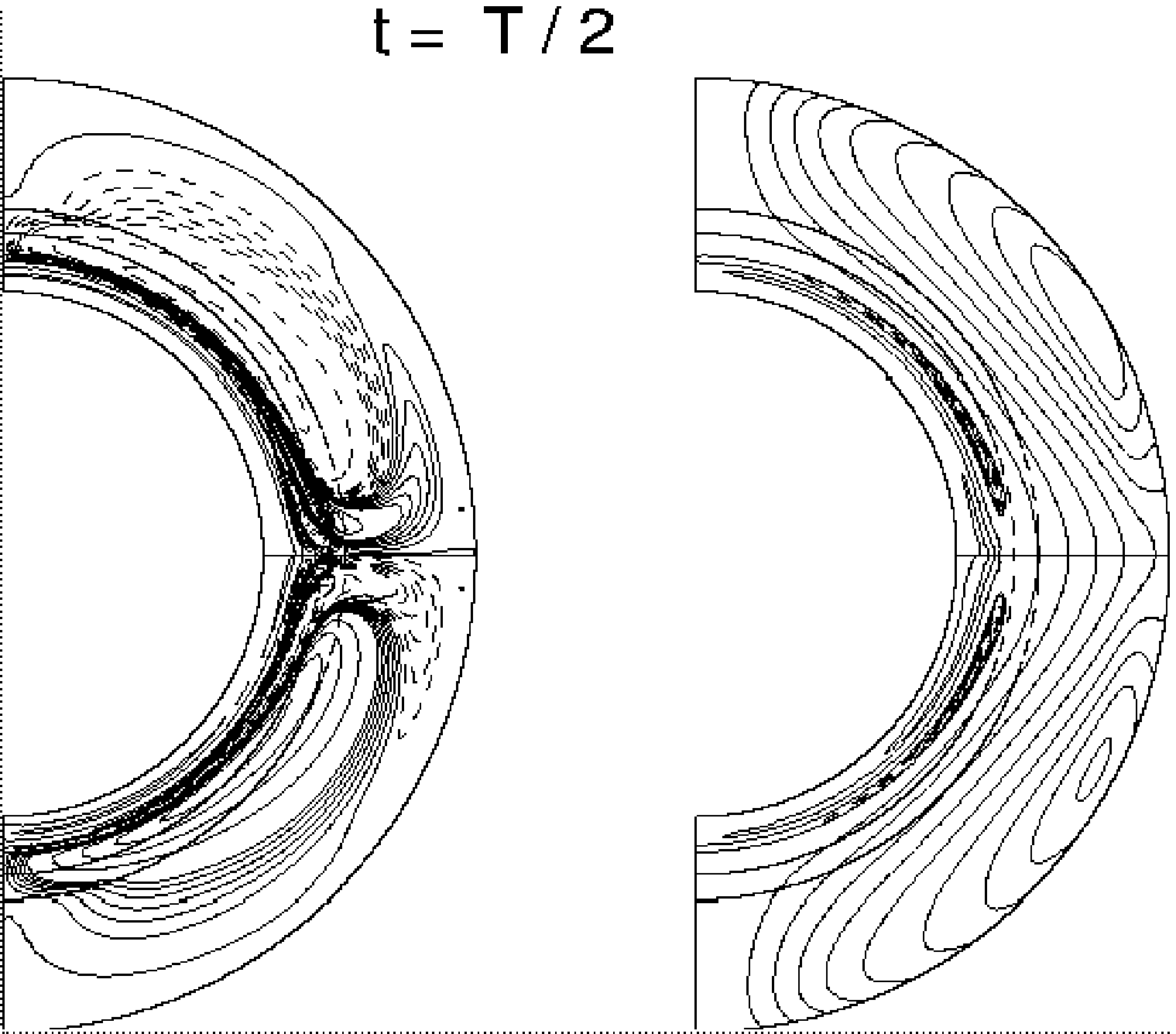}}
\caption{\label{fig:dipsnap} Four snapshots of the toroidal field contours (left panel), and the
poloidal field lines (right panel) separated by 1/8$^{\rm {th}}$ of the dynamo period $T$. 
The case $t = 0$ is shown in Fig.~\ref{fig:snap0}. The line styles are same as in 
Fig~\ref{fig:snap0}.}
\end{figure}

\subsection{The effect of velocity quenching}

One important question is whether the equatorward meridional circulation 
at the base of SCZ should be able to
advect the strong toroidal field, working against the magnetic tension.
It has been argued by Choudhuri (2003b) that this should be possible if
the strong toroidal field is highly intermittent.  However, Choudhuri (2003b)
concluded that the meridional flow may be barely strong enough to advect
the toroidal field.  If $B$ becomes larger than some critical value, then
the meridional flow may not be able to carry it.
We try to capture this effect by checking at intervals of 10 days if the toroidal
field exceeds a critical value of 1.5$\times$10$^{5}$ G in 
a region of thickness 0.11$R_{\odot}$ below a depth of 
0.73$R_{\odot}$. Whenever $B$ is larger than this critical value, 
we reduce the velocity at that grid 
point by a factor of eight. A butterfly diagram similar to Fig.~\ref{fig:dipradbfly}
is produced in this case and is shown in Fig.~\ref{fig:qnch}.  On comparing
Fig.~\ref{fig:dipradbfly} and Fig.~\ref{fig:qnch}, we find that this velocity 
quenching improves the
appearance of the butterfly diagram.  We saw in Fig.~\ref{fig:dipradbfly} that eruptions for
a new half-cycle began at high latitudes before the eruptions at low latitudes
stopped for the previous half-cycle, leading to a tail-like attachment in the
butterfly diagram.  We see in Fig.~\ref{fig:qnch} that this is gone and a new half-cycle 
begins at the high latitude at about the time when the old half-cycle dies
off at the low latitude---in conformity with observational data (see Fig.~6.2
of Choudhuri 2003a).

\begin{figure}[h!]
\label{fig:qnch}
\centerline{\includegraphics[width=10cm,height=8cm]{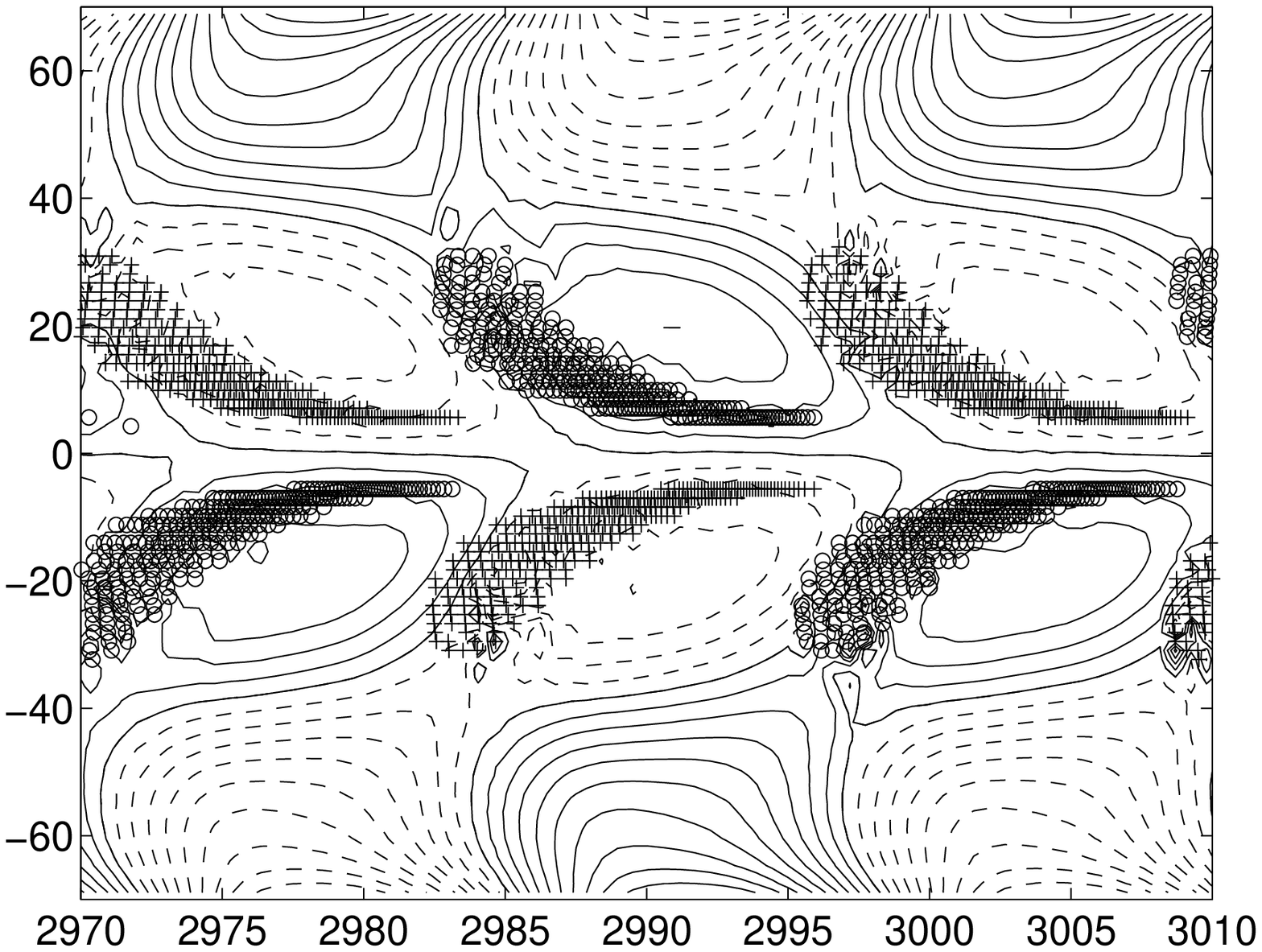}}
\caption{\label{fig:qnch} Theoretical butterfly diagram of eruptions for the case presented in
\S 4.1. The various conventions used are the same as in Fig~\ref{fig:dipradbfly}.}
\end{figure}

\section{Conclusion}

We show that a CDSD model with an $\alpha$ effect concentrated near the 
surface and a meridional circulation penetrating below the tachocline
provides a satisfactory explanation for various aspects of the solar cycle.
The helioseismically determined differential rotation is 
strongest at high latitudes within the tachocline
and there is little doubt that the strong toroidal field should be produced
there, to be advected by the meridional flow to low latitudes where it erupts.
This view, which was first put forth by Nandy \& Choudhuri (2002), is a
departure from the traditional viewpoint that the toroidal field is produced
basically at the same latitude where it erupts.  Ironically, this new
viewpoint makes the original motivation of Choudhuri et al.\ (1995) in
introducing the meridional circulation somewhat redundant.  A standard
result of dynamo theory (without any flow in the meridional plane), which
was first derived by Parker (1955), is that the product of $\alpha$
and {\diff} should be negative in the northern hemisphere, to ensure
the equatorward propagation of the dynamo wave (see, for example, Choudhuri
1998, \S16.5). The tilts of bipolar regions on the solar surface suggest that $\alpha$
should be positive in the northern hemisphere, whereas helioseismology
found {\diff} also to be positive in the lower latitudes.  If the 
strong toroidal field is produced within the tachocline at low latitudes
where it erupts,
then the simple sign rule would suggest a poleward propagation.  Choudhuri
et al.\ (1995) introduced the meridional circulation primarily to overcome
this tendency of poleward propagation, forcing the dynamo wave to propagate
equatorward.  If the toroidal field is produced at high latitudes where
{\diff} is negative, then the dynamo wave should propagate poleward
even in the absence of meridional circulation.  Although the original motivation
of Choudhuri et al.\ (1995) in introducing the meridional circulation may
no longer be so relevant, it has become increasingly clear in the last few
years that the meridional circulation plays a crucial role in the solar 
dynamo.  There are indications that the meridional circulation may actually
be the time-keeper of the solar cycle (Hathaway et al.\ 2003). We hope that
within the next few years helioseismology will discover the equatorward
return flow of meridional circulation and may even be able to establish if
this return flow really penetrates below the tachocline, as required by
us.

Dikpati \& Gilman (2001) and Bonanno et al.\ (2002) had earlier argued
that a pure Babcock--Leighton dynamo with $\alpha$ concentrated near the
surface may not give the correct dipolar parity.  We have clearly demonstrated
that this is not the case.  If the poloidal field has sufficient diffusivity
to get coupled across the equator, whereas the toroidal field is not able
to diffuse across the equator (since turbulent diffusion is suppressed for
the strong toroidal field), then we find that the dipolar mode is preferred.
We saw in \S3.3 that an additional $\alpha$ effect in the interior of SCZ
would help in establishing dipolar parity.  However, in view of the fact
that our knowledge about such an $\alpha$ effect is very uncertain, we felt
that it is first necessary to study pure Babcock--Leighton dynamo models
with $\alpha$ effect concentrated near the surface alone.  Accordingly, we
have taken such a dynamo model which gives the correct dipolar parity as
our standard model.  We have shown in \S4 that this standard model explains many
aspects of observational data very well.  We are right now exploring whether
this standard model can explain some other aspects of observational data not
discussed by us here.  For example, active regions in the northern hemisphere
are known to have a preferred negative helicity.  A theoretical explanation
for this has been provided by Choudhuri (2003b).  Our preliminary investigations based on the idea of Choudhuri (2003b)
show that our standard model presented in this paper would give the right type of helicity.  We are
now carrying out more detailed calculations, which will be reported in a
forthcoming paper. We may mention that our code can be used for other 
MHD calculations besides the solar dynamo problem.  A modified version of the
code has been used to study the evolution of magnetic fields in
neutron stars (Choudhuri \& Konar 2002; Konar \& Choudhuri 2004).

\bigskip

{\em Acknowledgements.}  Most of our calculations were carried out on {\it{San\'khya}}, 
the parallel cluster computer in Centre for High Energy Physics, Indian Institute
of Science. D. N. acknowledges financial support from NASA through SR\&T grant NAG5-11873.

\newcommand{\sop}{Solar Physics}
\newcommand{\gafd}{Geophys.\ Astrophys.\ Fluid Dyn.}
\section*{References}
Babcock, H.W. 1961, ApJ, 133, 572.\\
Bonanno, A., Elstner, D, R\"udiger, G., \& Belvedere, G., 2002, A\&A, 390, 673\\
Braun, D.C., \& Fan, Y. 1998, ApJ, 508, L105.\\
Caligari, P., Moreno-Insertis, F., \& Sch\"ussler, M. 1995, ApJ, 441,
886.\\
Charbonneau, P., Christensen-Dalsgaard, J., Henning, R., et al., 1999, ApJ, 527, 445\\
Choudhuri, A.R. 1989, \sop, 123, 217\\
Choudhuri, A.R. 1998, The Physics of Fluids and Plasmas: An Introduction for
Astrophysicists, Cambridge University Press, Cambridge.\\
Choudhuri, A.R., 2003a, in {Dynamic Sun}, ed B.N. Dwivedi, Cambridge University Press, 103.\\
Choudhuri, A.R., 2003b, Solar Phys, 215, 31 \\
Choudhuri, A.R., \& Dikpati, M. 1999, \sop, 184, 61\\
Choudhuri, A.R., \& Gilman, P.A. 1987, ApJ, 316, 788\\
Choudhuri, A.R., \& Konar, S., 2002, MNRAS, 332, 933 \\ 
Choudhuri, A.R., Sch\"ussler M., \& Dikpati M. 1995, A\&A, 303, L29\\
Dikpati, M., \& Charbonneau, P. 1999, ApJ, 518, 508\\
Dikpati, M., \& Choudhuri, A.R. 1994, A\&A, 291, 975\\
Dikpati, M., \& Choudhuri, A.R. 1995, \sop, 161, 9\\
Dikpati, M., \& Gilman, P.A. 2001, ApJ, 559, 428\\
D'Silva, S., \& Choudhuri, A.R. 1993, A\&A, 272, 621\\
D'Silva, S., \& Howard, R.F., 1993, \sop,  148, 1\\
Durney, B.R. 1995, \sop, 160, 213\\
Durney, B.R. 1996, \sop, 160, 231\\
Durney, B.R. 1997, ApJ, 486, 1065\\
Fan, Y., Fisher, G.H., \& DeLuca, E.E. 1993, ApJ, 405, 390\\
Ferriz-Mas, A., Schmitt, D., \& Sch\"ussler, M., 1994, ApJ, 289, 949\\
Giles, P.M., Duvall, T. L., Jr., Kosovichev, A. G., \& Scherrer, P. H., 1997, 
Nature, 390, 52\\
Gilman, P.A. 1983, ApJS, 53, 243\\
Gilman, P.A. 1986, in { Physics of the Sun}, Vol 1, 95-160, Dordrecht, D. Reidel 
Publishing Co. \\
Hathaway, D. H., Nandi, D., \& Wilson, R. M., Reichmann, E.J., 2003, 
Astrophys. J., 589, 665\\
Konar, S., \& Choudhuri A.R., 2004, MNRAS, 348, 661\\
K\"uker, M., R\"udiger, G. \& Schultz, M. 2001, A\&A, 374, 301\\
Leighton, R.B. 1969, ApJ, 156, 1\\
Longcope, D. \& Choudhuri, A.R., 2002, \sop, 205, 63\\
Miesch, M. S., Elliott, J. R., Toomre, J.,  et al. 2000, ApJ, 532, 593\\
Nandy, D., 2002 Ap\&SS, 282, 209 \\
Nandy, D., 2003, { Modelling the Solar Magnetic Cycle}, PhD Thesis, Dept of Physics, Indian Institute of Science\\
Nandy, D., \& Choudhuri, A.R. 2000, J. Astrophys. Astron., 21, 381\\
Nandy, D., \& Choudhuri, A.R. 2001, ApJ, 551, 576\\
Nandy, D. \& Choudhuri, A.R., 2002, Science, 296, 1671\\
Parker, E.N. 1955, ApJ, 122, 293\\
Schmitt, D., \& Sch\"ussler, M., 1989, A\&A, 223, 343\\
Schou, J., Antia, H. M., Basu, S., et al, 1998, ApJ, 505, 390\\
Steenbeck, M., \& Krause, F., R\"adler, K.H., 1966, Z. Naturforsch., vol-21a, 369\\
Wang, Y.-M., Nash, A.G., \& Sheeley, N.R. 1989a, ApJ, 347, 529\\
Wang, Y.-M., Nash, A.G., \& Sheeley, N.R. 1989b, Science, 245, 712\\
Wang, Y.-M., Sheeley, N.R., \& Nash, A.G. 1991, ApJ, 383, 431\\

\end{document}